\documentclass[twocolumn,preprintnumbers,amsmath,amssymb]{revtex4}  

\usepackage{graphicx}
\usepackage{dcolumn}
\usepackage{bm}
\usepackage{mathrsfs} 

\begin{document}

\title{Possible two-component pairings in electron-doped Bi$_{2}$Se$_{3}$ based on a tight-binding model}

\author{Lei Hao$^1$ and C. S. Ting$^2$}
\address{$^1$School of Physics, Southeast University, Nanjing 211189, China  \notag \\
$^2$Department of Physics and Texas Center for Superconductivity, University of Houston, Houston, Texas 77204, USA}

\date{\today}

\begin{abstract}
Recent experiments show the spontaneous breaking of rotational symmetry in the superconducting topological insulators M$_{x}$Bi$_{2}$Se$_{3}$ (M represents Cu, Sr, or Nd), suggesting that the pairing belongs to a two-dimensional representation of the $D_{3d}$ symmetry group of the crystal. Motivated by these progresses, we construct an exhaustive list of possible two-component pairings of the M$_{x}$Bi$_{2}$Se$_{3}$ superconductors, both for the odd-parity $E_{u}$ representation and for the even-parity $E_{g}$ representation. Starting from a tight-binding model for the normal phase of Bi$_{2}$Se$_{3}$ and M$_{x}$Bi$_{2}$Se$_{3}$, we firstly construct the pairing channels in the spin-orbital basis, up to second-nearest-neighbor pairing correlations in the basal plane. We then infer the properties of these pairings by transforming them to the band (pseudospin) basis for the conduction band. A comparison with the key experimental consensuses on M$_{x}$Bi$_{2}$Se$_{3}$ superconductors shows that the true pairings should also be multichannel. Besides a nematic and time-reversal symmetric pairing combination, the other pairings that we have identified are chiral and nematic at the same time, which may be nonunitary and have a spontaneous magnetization. A complementary set of experiments are proposed to identify the true pairing symmetries of these superconductors and their evolution with the doping concentration $x$.
\end{abstract}


\maketitle

\section{Introduction}

The nature of the pairing of superconducting topological insulators, to be abbreviated as M$_{x}$Bi$_{2}$Se$_{3}$ with M representing a metallic element that might be Cu, Sr or Nd, has been mysterious since their discovery \cite{hor10,wray10,liu15,shruti15,han15,qiu15}. Various experiments made conflicting implications, alluding the pairing to be topologically nontrivial \cite{sasaki11,kirzhner12,chen12,bay12,das11} or trivial \cite{levy13,peng13}. Recently, a series of new experiments have shown convincingly that the pairing of this family of superconductors is unconventional (see Yonezawa \cite{yonezawa18} for a recent review). For Cu$_{x}$Bi$_{2}$Se$_{3}$ \cite{matano16,yonezawa17} and Sr$_{x}$Bi$_{2}$Se$_{3}$ \cite{du17,pan16}, more than one experiments have revealed that the superconducting state breaks the three-fold rotational symmetry of the normal phase to two-fold rotational symmetry, which is possible only if the pairing belongs to a two-dimensional representation of the underlying $D_{3d}$ point group. The pairing was thereby called nematic superconductivity \cite{fu14}. For Nd$_{x}$Bi$_{2}$Se$_{3}$, in addition to the broken three-fold in-plane rotational symmetry \cite{asaba17,shen17}, the time-reversal symmetry appears to be also broken \cite{qiu15}.

Multiple theoretical analysis have been made to account for these new experimental findings \cite{fu16,venderbos16a,venderbos16b,nagai16,yuan17,chirolli17,wu17,wu17b,zyuzin17,hao17}, which focus mostly on the odd-parity $E_{u}$ representation of the $D_{3d}$ point group. While the studies can account for the qualitative features of various experiments, a satisfactory explanation of all crucial experimental features in terms of known pairings appears to be difficult. For example, while recent experiments deny the presence of in-gap states on the surface of superconducting Cu$_{x}$Bi$_{2}$Se$_{3}$ \cite{levy13,peng13}, the most studied $E_{u}$ pairing is shown to have robust low-energy surface states \cite{hao17}. In addition, the proposed $E_{u}$ pairing has not been stabilized as the leading pairing instability in any theoretical calculations based on a microscopic pairing mechanism, such as the electron-phonon interaction \cite{fu10,wan14,brydon14,zhang15} or the electron-electron interaction \cite{hao13}. In view of the difficulty in first-principles predictions of the pairing symmetry on one hand and the extensive experimental observations accumulated up to date on the other hand, a promising approach is to construct an exhaustive list of possible two-component pairings and compare them with the available experimental consensuses. From this comparison, we may see to what extent the existing experiments have constrained the pairing symmetry, and what further experiments are necessary to figure out definitely the genuine pairing symmetries of the M$_{x}$Bi$_{2}$Se$_{3}$ superconductors.

Motivated by the above considerations, we construct in this work a complete list of pairing channels belonging to the two-dimensional irreducible representations of the $D_{3d}$ point group. Starting from a tight-binding model for Bi$_{2}$Se$_{3}$ and the normal phase of M$_{x}$Bi$_{2}$Se$_{3}$, we construct pairings belonging to the odd-parity $E_{u}$ representation and pairings in the even-parity $E_{g}$ representation. In consistency with the tight-binding model which is up to second-nearest-neighbor (2NN) in-plane hoppings, the constructed pairing channels are restricted to 2NN in-plane pairing correlations. By transforming from the spin-orbital basis to the band (pseudospin) basis and retaining only the conduction band which contributes to the Fermi surface, we analyze the general properties of various interesting and typical pairing channels.

After a comprehensive review over the experimental consensuses on these superconductors, including their bulk spectrum, surface spectrum, and magnetic properties, we infer the constraints of these experiments on the pairing symmetry. The major conclusion of the comparison is that the pairing has to be multichannel, in addition to having two components. A purely nematic pairing combination in the $E_{u}$ representation can give a fully-gapped and two-fold symmetric bulk spectrum, fully-gapped surface spectrum, and two-fold symmetric electronic spin susceptibility. In addition, we find several chiral and nematic pairing combinations that can explain the key experimental results, in both the $E_{g}$ representation and the $E_{u}$ representation. The chiral and nematic pairing of the $E_{u}$ representation, besides breaking the time-reversal symmetry, may also be nonunitary and having a spontaneous magnetization. We finally discuss the implications of the present work to the future experiments, which are highly desirable to determine the true pairing symmetry of the M$_{x}$Bi$_{2}$Se$_{3}$ superconductors.

\section{Model in the spin-orbital and pseudospin basis}

The low-energy band structures of Bi$_{2}$Se$_{3}$ and M$_{x}$Bi$_{2}$Se$_{3}$ (M denotes Cu, Sr, or Nd) can be described by the following two-orbital tight-binding model, defined on a quasi-two-dimensional hexagonal lattice \cite{zhang09,fu09,liu10,hao13,hao17},
\begin{eqnarray}
H_{0}(\mathbf{k})&=&\epsilon(\mathbf{k})I_{4}+M(\mathbf{k})\Gamma_5+B_{0}c_{z}(\mathbf{k})\Gamma_{4}
+A_{0}[c_{y}(\mathbf{k})\Gamma_{1}   \notag \\
&&-c_{x}(\mathbf{k})\Gamma_{2}]+R_{1}d_{1}(\mathbf{k})\Gamma_{3}+R_{2}d_{2}(\mathbf{k})\Gamma_{4}.
\end{eqnarray}
The basis operator is taken as $\phi^{\dagger}_{\mathbf{k}}=[a^{\dagger}_{\mathbf{k}\uparrow}, a^{\dagger}_{\mathbf{k}\downarrow}, b^{\dagger}_{\mathbf{k}\uparrow}, b^{\dagger}_{\mathbf{k}\downarrow}]$, where the $a$ and $b$ orbitals separately correspond to the $p_{z}$ orbitals of the top and bottom Se layers of the Bi$_{2}$Se$_{3}$ quintuple units, with a certain amount of hybridization with the $p_{z}$ orbitals of the neighboring Bi layers \cite{zhang09,fu09,liu10}. $I_4$ is the $4\times4$ unit matrix. $\Gamma_1=\sigma_{3}\otimes s_{1}$, $\Gamma_2=\sigma_{3}\otimes s_{2}$, $\Gamma_3=\sigma_{3}\otimes s_{3}$,
$\Gamma_{4}=-\sigma_{2}\otimes s_{0}$, and $\Gamma_{5}=\sigma_{1}\otimes s_{0}$ \cite{wang10,hao11,fu09,fu10,liu10,zhang09,hao13}.
$s_{i}$ and $\sigma_{i}$ ($i=1,2,3$) are Pauli matrices for the spin and orbital degrees of freedom, $s_{0}$ and $\sigma_{0}$ are the corresponding unit matrices. With the parity operator $P=\sigma_{1}\otimes s_{0}$, it is easy to verify that the model has the inversion symmetry $PH_{0}(\mathbf{k})P^{-1}=H_{0}(-\mathbf{k})$.

The above model was obtained previously \cite{hao13} based on symmetry analysis and comparison with a $\mathbf{k}\cdot\mathbf{p}$ model defined near $k_{x}=k_{y}=k_{z}=0$ \cite{liu10}. The lattice of Bi$_2$Se$_3$ and M$_{x}$Bi$_2$Se$_3$, which belong to the $D_{3d}^{5}$ space group, is mapped to a hexagonal lattice in the tight-binding model. The in-plane (labeled as the $xy$ plane) and out-of-plane (labeled as the $z$ direction) lattice parameters, $a$ and $c$, are taken as $a$=4.14 \text{\AA} and $3c$=28.64 \text{\AA} \cite{acparameters}. $\epsilon(\mathbf{k})=C_{0}+2C_{1}[1-\cos(\mathbf{k}\cdot\boldsymbol{\delta}_{4})]
+\frac{4}{3}C_{2}[3-\cos(\mathbf{k}\cdot\boldsymbol{\delta}_{1})-\cos(\mathbf{k}\cdot\boldsymbol{\delta}_{2})
-\cos(\mathbf{k}\cdot\boldsymbol{\delta}_{3})]$. $M(\mathbf{k})$ is obtained from $\epsilon(\mathbf{k})$
by making the substitutions $C_{i}\rightarrow M_{i} (i=0,1,2)$. $c_{x}(\mathbf{k})=\frac{1}{\sqrt{3}}
[\sin(\mathbf{k}\cdot\boldsymbol{\delta}_{1})-\sin(\mathbf{k}\cdot\boldsymbol{\delta}_{2})]$,
$c_{y}(\mathbf{k})=\frac{1}{3}[\sin(\mathbf{k}\cdot\boldsymbol{\delta}_{1})+\sin(\mathbf{k}\cdot\boldsymbol{\delta}_{2})
-2\sin(\mathbf{k}\cdot\boldsymbol{\delta}_{3})]$, and $c_{z}(\mathbf{k})=\sin(\mathbf{k}\cdot\boldsymbol{\delta}_{4})$.
Finally, $d_{1}(\mathbf{k})=-\frac{8}{3\sqrt{3}}[\sin(\mathbf{k}\cdot\mathbf{a}_{1})+\sin(\mathbf{k}\cdot\mathbf{a}_{2})
+\sin(\mathbf{k}\cdot\mathbf{a}_{3})]$ and $d_{2}(\mathbf{k})=-8[\sin(\mathbf{k}\cdot\boldsymbol{\delta}_{1})
+\sin(\mathbf{k}\cdot\boldsymbol{\delta}_{2})+\sin(\mathbf{k}\cdot\boldsymbol{\delta}_{3})]$. Here, the four NN bond vectors of the hexagonal lattice are $\boldsymbol{\delta}_{1}=(\frac{\sqrt{3}}{2}a, \frac{1}{2}a, 0)$,
$\boldsymbol{\delta}_{2}=(-\frac{\sqrt{3}}{2}a$, $\frac{1}{2}a, 0)$, $\boldsymbol{\delta}_{3}=(0, -a, 0)$, and $\boldsymbol{\delta}_{4}=(0, 0, c)$.
The three in-plane 2NN bond vectors in $d_{1}(\mathbf{k})$ are $\mathbf{a}_{1}=\boldsymbol{\delta}_{1}-\boldsymbol{\delta}_{2}$,
$\mathbf{a}_{2}=\boldsymbol{\delta}_{2}-\boldsymbol{\delta}_{3}$, and $\mathbf{a}_{3}=\boldsymbol{\delta}_{3}-\boldsymbol{\delta}_{1}$. The last and second last terms of $H_{0}(\mathbf{k})$ induce hexagonal warping of the Fermi surface and the topological surface states \cite{fu09,liu10}. We mention in passing that Nd$_{x}$Bi$_{2}$Se$_{3}$ was reported to have multiple Fermi surfaces, with possible contributions from the $d$ orbitals of the Nd dopants \cite{lawson16}. We will neglect this complexity and work with the above model for all three superconductors \cite{yuan17,chirolli17}.

The dopants of M$_{x}$Bi$_{2}$Se$_{3}$ (M is Cu, Sr, or Nd) dope electrons to the system, so that only the conduction band contributes to the Fermi surface. In studying the bulk properties of a superconductor, it is easier to work with the band (pseudospin) basis and retain only the states of the conduction band \cite{yip13,yip16,zocher13,hashimoto13,nagai14,takami14,hao14}. For this purpose, we turn to the pseudospin basis for the conduction band. Because M$_{x}$Bi$_{2}$Se$_{3}$ in the normal state has both inversion symmetry and time-reversal symmetry, the conduction band is twofold degenerate (the Kramers degeneracy) at each wave vector $\mathbf{k}$. The operator for the inversion symmetry is the parity operator $P=\sigma_{1}\otimes s_{0}$. The time reversal operator is taken as $T=-i\sigma_{0}\otimes s_{2}K$, where $K$ denotes the complex conjugation. We define the pseudospin basis for the conduction band states on the northern hemisphere (i.e., $k_{z}>0$) of the three-dimensional Brillouin zone (BZ) as
\begin{equation}
[|\mathbf{k},\alpha\rangle,|\mathbf{k},\beta\rangle]=[|\mathbf{k},\alpha'\rangle,|\mathbf{k},\beta'\rangle]u_{\mathbf{k}},
\end{equation}
where $\alpha$ and $\beta$ are the two pseudospin degrees of freedom. The Kramers degeneracy relates the two bases via $|\mathbf{k},\beta\rangle=PT|\mathbf{k},\alpha\rangle$. The two auxiliary bases are defined as \cite{hao17}
\begin{equation}
|\mathbf{k},\alpha'\rangle=\frac{1}{\tilde{D}_{\mathbf{k}}N_{\mathbf{k}}}\begin{pmatrix} \tilde{E}_{\mathbf{k}} \\ \tilde{M}_{-}(\mathbf{k}) \end{pmatrix} \begin{pmatrix} A_{0}c_{+}(\mathbf{k}) \\ D_{-}(\mathbf{k}) \end{pmatrix},
\end{equation}
and
\begin{equation}
|\mathbf{k},\beta'\rangle=PT|\mathbf{k},\alpha'\rangle=\frac{1}{\tilde{D}_{\mathbf{k}}N_{\mathbf{k}}}\begin{pmatrix} \tilde{M}_{+}(\mathbf{k})  \\ \tilde{E}_{\mathbf{k}} \end{pmatrix} \begin{pmatrix} -D_{-}(\mathbf{k})  \\ A_{0}c_{-}(\mathbf{k})   \end{pmatrix},
\end{equation}
where the first and second two-component vectors are separately spinors in the subspaces of the original orbital and spin degrees of freedom. The unitary matrix connecting the two basis sets is \cite{hao17}
\begin{equation}
u_{\mathbf{k}}=\begin{pmatrix} ie^{i(\varphi_{\mathbf{k}}+\phi_{\mathbf{k}})}\cos\frac{\theta_{\mathbf{k}}}{2}  &  -e^{i\phi_{\mathbf{k}}}\sin\frac{\theta_{\mathbf{k}}}{2}  \\ e^{-i\phi_{\mathbf{k}}}\sin\frac{\theta_{\mathbf{k}}}{2} & -ie^{-i(\varphi_{\mathbf{k}}+\phi_{\mathbf{k}})}\cos\frac{\theta_{\mathbf{k}}}{2}  \end{pmatrix}.
\end{equation}
For notational simplicity, we have introduced the following abbreviations in Eqs. (3)-(5): $c_{\pm}(\mathbf{k})=c_{y}(\mathbf{k})\pm i c_{x}(\mathbf{k})$, $\tilde{M}_{\pm}(\mathbf{k})=M(\mathbf{k})\pm i[B_{0}c_{z}(\mathbf{k})+R_{2}d_{2}(\mathbf{k})]$, $D_{\mathbf{k}}=\sqrt{A_{0}^{2}[c_{x}^{2}(\mathbf{k})+c_{y}^{2}(\mathbf{k})]+R_{1}^{2}d_{1}^{2}(\mathbf{k})}$, $E_{\mathbf{k}}=\sqrt{|\tilde{M}_{\pm}(\mathbf{k})|^{2}+D_{\mathbf{k}}^{2}}$, $\tilde{E}_{\mathbf{k}}=E_{\mathbf{k}}+D_{\mathbf{k}}$, $N_{\mathbf{k}}=\sqrt{2E_{\mathbf{k}}\tilde{E}_{\mathbf{k}}}$, $D_{\pm}(\mathbf{k})=D_{\mathbf{k}}\pm R_{1}d_{1}(\mathbf{k})$, $\tilde{D}_{\mathbf{k}}=\sqrt{2D_{\mathbf{k}}D_{-}(\mathbf{k})}$, and
\begin{equation}
c_{+}(\mathbf{k})=i\sqrt{c_{x}^{2}(\mathbf{k})+c_{y}^{2}(\mathbf{k})}e^{-i\varphi_{\mathbf{k}}}=ic(\mathbf{k})e^{-i\varphi_{\mathbf{k}}},
\end{equation}
\begin{equation}
W_{\mathbf{k}}=\frac{\tilde{E}_{\mathbf{k}}+\tilde{M}_{+}(\mathbf{k})}{\sqrt{2}N_{\mathbf{k}}}=|W_{\mathbf{k}}|e^{i\phi_{\mathbf{k}}},
\end{equation}
\begin{equation}
R_{1}d_{1}(\mathbf{k})+iA_{0}c(\mathbf{k})=D_{\mathbf{k}}e^{i\theta_{\mathbf{k}}}.
\end{equation}

The above formulae define the pseudospin basis for the conduction band states on the northern hemisphere of the BZ ($k_{z}>0$). For conduction band states on the southern hemisphere (i.e., $k_{z}<0$), the pseudospin basis are related to the pseudospin basis for states on the northern hemisphere by the symmetry operations: $|\mathbf{k},\alpha\rangle=P|-\mathbf{k},\alpha\rangle=-T|-\mathbf{k},\beta\rangle$ and $|\mathbf{k},\beta\rangle=P|-\mathbf{k},\beta\rangle=T|-\mathbf{k},\alpha\rangle$.
We introduce the new Pauli matrices $\varrho_{i}$ ($i=1,2,3$) and the corresponding unit matrix $\varrho_{0}$ in the subspace of the two pseudospin bases. The reduced model containing only states of the conduction band is simply
\begin{equation}
h_{0}(\mathbf{k})=E(\mathbf{k})\varrho_{0},
\end{equation}
where the dispersion of the conduction band is $E(\mathbf{k})=\epsilon(\mathbf{k})+E_{\mathbf{k}}$.

\section{Lists and general properties of two-component pairings}

We now construct the full lists of the basis functions for the $E_{g}$ and $E_{u}$ representations of the $D_{3d}$ group, up to 2NN in-plane pairing correlations, consistent with the tight-binding model which is up to 2NN in-plane hopping terms \cite{hao13}. Corresponding to the two basis of the model defined in the previous section, there are two ways of classifying the possible pairing channels in M$_{x}$Bi$_{2}$Se$_{3}$. The first approach focuses on the low-energy states close to the Fermi surface \cite{venderbos16a,venderbos16b,wu17}. For the $x>0$ case of all three superconductors, the Fermi surface consists of states in the conduction band. Then we can neglect the valence band from our full model and work with a reduced model with only the states in the conduction band. In the second approach, we work with the full two-orbital model and construct the basis functions in the spin-orbital basis \cite{fu10,yuan17,chirolli17}. If we are interested only in the low-energy properties of the M$_{x}$Bi$_{2}$Se$_{3}$ superconductor in the bulk, or if the normal phase is topologically trivial, the two approaches give essentially the same results. However, if we are also interested in the topological aspect of the system inherited from the topologically nontrivial normal phase, such as the coexistence of the topological surface states with the Fermi surface \cite{wray10,lahoud13,tanaka12}, then it is advantageous, if not imperative, to work with the second approach.

We will first construct in the spin-orbital basis the full lists of pairing channels in both the $E_{g}$ and the $E_{u}$ representations, up to 2NN in-plane pairing correlations. Basis functions for the irreducible representations of the $D_{3d}$ symmetry group can be constructed in terms of the $\Gamma$ matrices or the symmetrized Fourier functions. Here, we define the `symmetrized Fourier functions' as linear combinations of the trigonometric functions $\cos(\mathbf{k}\cdot\mathbf{l})$ and $\sin(\mathbf{k}\cdot\mathbf{l})$, where $\mathbf{k}$ is the wave vector and $\mathbf{l}$ represents an NN or 2NN bond vectors defined in section II. Full lists of these basis functions exist in previous works \cite{liu10,hao13}. To be self-contained, we include them in Table I and Table II. For each representation, there are two sets of basis functions in terms of the $\Gamma$ matrices. Up to a constant number of unit module, the two basis sets differ by a factor of $\Gamma_{5}$. This is easy to understand from the fact that $\Gamma_{5}$ belongs to the $A_{1g}$ representation, which respects the full symmetry of the crystal and maps an existing basis set to a new basis set belonging to the same representation. For the $E_{g}$ and $E_{u}$ representations, the two basis sets in Table I transform in the same manner under the $D_{3d}$ group \cite{hao13}.
The symmetrized Fourier functions in Table II and their expansions in the limit of small in-plane wave vectors (i.e., $k_{x}a\simeq0$ and $k_{y}a\simeq0$) are
\begin{eqnarray}
\varphi_{0}(\mathbf{k})&=&\frac{1}{3}[\cos\mathbf{k}\cdot\boldsymbol{\delta}_{1} +\cos\mathbf{k}\cdot\boldsymbol{\delta}_{2}+\cos\mathbf{k}\cdot\boldsymbol{\delta}_{3}]   \notag \\
&\simeq& 1-\frac{1}{6}(k_{x}^{2}+k_{y}^{2})a^{2},
\end{eqnarray}
\begin{equation}
\varphi_{1}(\mathbf{k})=\frac{1}{2}[\cos\mathbf{k}\cdot\boldsymbol{\delta}_{1}-\cos\mathbf{k}\cdot\boldsymbol{\delta}_{2}]  \simeq-\frac{\sqrt{3}}{4}k_{x}k_{y}a^{2},
\end{equation}
\begin{eqnarray}
\varphi_{2}(\mathbf{k})&=&\frac{[\cos\mathbf{k}\cdot\boldsymbol{\delta}_{1} +\cos\mathbf{k}\cdot\boldsymbol{\delta}_{2}-2\cos\mathbf{k}\cdot\boldsymbol{\delta}_{3}]}{2\sqrt{3}}   \notag \\
&\simeq&-\frac{\sqrt{3}}{8}(k_{x}^{2}-k_{y}^{2})a^{2},
\end{eqnarray}
\begin{equation}
\varphi_{3}(\mathbf{k})=d_{1}(\mathbf{k})\simeq(k_{x}^{3}-3k_{x}k_{y}^{2})a^{3} =\frac{1}{2}(k_{+}^{3}+k_{-}^{3})a^{3},
\end{equation}
\begin{equation}
\varphi_{4}(\mathbf{k})=d_{2}(\mathbf{k})\simeq(3k_{y}k_{x}^{2}-k_{y}^{3})a^{3} =\frac{1}{2i}(k_{+}^{3}-k_{-}^{3})a^{3},
\end{equation}
\begin{equation}
\varphi_{5}(\mathbf{k})=c_{x}(\mathbf{k})\simeq k_{x}a,
\end{equation}
and
\begin{equation}
\varphi_{6}(\mathbf{k})=c_{y}(\mathbf{k})\simeq k_{y}a.
\end{equation}
We have introduced the abbreviation $k_{\pm}=k_{x}\pm ik_{y}$. If we extend to include the  inter-quintuple-layer pairings, we can replace $\varphi_{0}(\mathbf{k})$ with $\varphi'_{0}(\mathbf{k})=\cos\mathbf{k}\cdot\boldsymbol{\delta}_{4}$ and replace $\varphi_{4}(\mathbf{k})$ with $\varphi'_{4}(\mathbf{k})=\sin\mathbf{k}\cdot\boldsymbol{\delta}_{4}$. However, we will focus on the intra-quintuple-layer pairings in this work.

\begin{table}[ht] 
\caption{Basis functions in terms of the $\Gamma$ matrices. The symbols in the brackets of the first column are another commonly used name for the corresponding representation \cite{liu10,hao13}. The semicolons in the second column separate different basis sets of the same representation.} \centering
\begin{tabular}{c c}
\hline\hline \\ [-1.5ex]
Representation \hspace{1cm} & Basis \\ [0.2ex]
\hline  \\ [-2ex]
$A_{1g}$($\tilde{\Gamma}_{1}^{+}$)  \hspace{1cm}  & $I_{4}$; $\Gamma_{5}$  \\
\hline  \\ [-2ex]
$A_{2g}$($\tilde{\Gamma}_{2}^{+}$)  \hspace{1cm}  & $\Gamma_{12}$; $\Gamma_{34}$  \\
\hline  \\ [-2ex]
$E_{g}$($\tilde{\Gamma}_{3}^{+}$)  \hspace{1cm}  & \{$\Gamma_{13}$,$\Gamma_{23}$\}; \{$\Gamma_{24}$,$\Gamma_{41}$\}  \\
\hline  \\ [-2ex]
$A_{1u}$($\tilde{\Gamma}_{1}^{-}$)  \hspace{1cm}  & $\Gamma_{3}$; $\Gamma_{35}$  \\
\hline  \\ [-2ex]
$A_{2u}$($\tilde{\Gamma}_{2}^{-}$)  \hspace{1cm}  & $\Gamma_{4}$; $\Gamma_{45}$  \\
\hline  \\ [-2ex]
$E_{u}$($\tilde{\Gamma}_{3}^{-}$)  \hspace{1cm}  & \{$\Gamma_{1}$,$\Gamma_{2}$\}; \{$\Gamma_{15}$,$\Gamma_{25}$\}  \\
\hline
\hline
\end{tabular}
\end{table}

\begin{table}[ht] 
\caption{Basis functions in terms of the symmetrized Fourier functions. The symmetrized Fourier functions are defined as linear combinations of $\cos(\mathbf{k}\cdot\mathbf{l})$ and $\sin(\mathbf{k}\cdot\mathbf{l})$, where $\mathbf{k}$ is the wave vector and $\mathbf{l}$ represents an NN or 2NN bond vectors defined below Eq.(1). The symbols in the brackets of the first column are another commonly used name for the corresponding representation \cite{liu10,hao13}. The semicolon in the second line of the second column separates different basis sets of the $A_{1g}$ representation.} \centering
\begin{tabular}{c c}
\hline\hline \\ [-1.5ex]
Representation \hspace{1cm} & Basis \\ [0.2ex]
\hline  \\ [-2ex]
$A_{1g}$($\tilde{\Gamma}_{1}^{+}$)  \hspace{1cm}  & $1$; $\varphi_{0}(\mathbf{k})$  \\
\hline  \\ [-2ex]
$A_{2g}$($\tilde{\Gamma}_{2}^{+}$)  \hspace{1cm}  &   none  \\
\hline  \\ [-2ex]
$E_{g}$($\tilde{\Gamma}_{3}^{+}$)  \hspace{1cm}  & \{$\varphi_{1}(\mathbf{k})$,$\varphi_{2}(\mathbf{k})$\}  \\
\hline  \\ [-2ex]
$A_{1u}$($\tilde{\Gamma}_{1}^{-}$)  \hspace{1cm}  & $\varphi_{3}(\mathbf{k})$  \\
\hline  \\ [-2ex]
$A_{2u}$($\tilde{\Gamma}_{2}^{-}$)  \hspace{1cm}  & $\varphi_{4}(\mathbf{k})$  \\
\hline  \\ [-2ex]
$E_{u}$($\tilde{\Gamma}_{3}^{-}$)   \hspace{1cm}  & \{-$\varphi_{6}(\mathbf{k})$,$\varphi_{5}(\mathbf{k})$\}  \\
\hline
\hline
\end{tabular}
\end{table}

By multiplying the basis functions in Table I and those in Table II, we can get various product representations of the $D_{3d}$ group. These product representations can be decomposed into the irreducible representations according to the group theory \cite{dresselhaus08}. For example, $A_{1u}\otimes E_{g}=E_{u}$ and $E_{g}\otimes E_{u}=A_{1u}+A_{2u}+E_{u}$. In such a manner, we can identify all the realizations of the possible irreducible representations. When taken as a part of the model Hamiltonian, they are subject to further constraints. If a term is taken as a part of the model for the electronic structures in the normal state, this term should belong to the $A_{1g}$ representation and has to be Hermitian \cite{liu10,hao13}. These constraints, together with the time-reversal symmetry of the materials, eliminate many combinations.

If a basis set is taken as the superconducting pairing term, then it has to satisfy the Fermi exchange statistics. Another important aspect about the symmetry of the pairing term is related to the peculiar transformation property of the pairing term under the time-reversal operation \cite{blount85,yip14}. In the spin-orbital basis $\phi^{\dagger}_{\mathbf{k}}$, the pairing term has the following general expression
\begin{equation}
\phi^{\dagger}_{\mathbf{k}}\underline{\Delta}(\mathbf{k})[\phi^{\dagger}_{-\mathbf{k}}]^{\text{T}}+\text{H.c.},
\end{equation}
where the superscript `T' means taking the transpose, and H.c. means the Hermitian conjugate of the first term. According to Blount \cite{blount85,yip14}, the time-reversed creation operator transforms under the symmetry operation just like the corresponding annihilation operator. That is,
\begin{equation}
T[\phi^{\dagger}_{\mathbf{k}}]^{\text{T}}=-i\sigma_{0}\otimes s_{2}[\phi^{\dagger}_{-\mathbf{k}}]^{\text{T}}
\end{equation}
transforms in the same manner as $\phi_{\mathbf{k}}$ under the action of the $D_{3d}$ symmetry group and the time-reversal operation \cite{blount85}. This means that, by transforming the creation operator part of the pairing term to
\begin{equation}
\phi^{\dagger}_{\mathbf{k}}\underline{\Delta}(\mathbf{k})[\phi^{\dagger}_{-\mathbf{k}}]^{\text{T}}
=\phi^{\dagger}_{\mathbf{k}}\underline{\Delta}(\mathbf{k})i\sigma_{0}\otimes s_{2}(-i\sigma_{0}\otimes s_{2})[\phi^{\dagger}_{-\mathbf{k}}]^{\text{T}},
\end{equation}
the matrix
\begin{equation}
\underline{\Delta}(\mathbf{k})i\sigma_{0}\otimes s_{2}
\end{equation}
has the same transformation property as the terms in the model for the normal state electronic structures. As a result, we can construct the basis functions according to the general procedure for the normal state \cite{dresselhaus08,liu10}. Since the obtained basis is of the form of Eq.(20), we multiply $-i\sigma_{0}\otimes s_{2}$ \emph{from the right} and get the basis functions of the pairing term. Then we single out from the results those obeying the Fermi exchange statistics, namely $\underline{\Delta}^{\text{T}}(-\mathbf{k})=-\underline{\Delta}(\mathbf{k})$. The above discussions correspond to taking the Nambu basis as $[\phi^{\dagger}_{\mathbf{k}},\phi^{\text{T}}_{-\mathbf{k}}]$. If we take the Nambu basis as $[\phi^{\dagger}_{\mathbf{k}},\phi^{\text{T}}_{-\mathbf{k}}(i\sigma_{0}\otimes s_{2})]$ instead, then the pairing term will be in the form of Eq.(20) spontaneously \cite{yip13}. Hereafter, we will stick to the first Nambu basis.

The resulting basis sets (up to 2NN in-plane pairing correlations) for the $E_{g}$ and $E_{u}$ representations are separately shown in Table III and Table IV. The new $\Gamma$-matrices in the tables with two subindices are defined as $\Gamma_{\mu\nu}=\frac{1}{2i}[\Gamma_{\mu},\Gamma_{\nu}]$, where both $\mu$ and $\nu$ run from $1$ to $5$. Explicitly, $\Gamma_{12}=\sigma_{0}\otimes s_{3}$, $\Gamma_{13}=-\sigma_{0}\otimes s_{2}$, $\Gamma_{14}=\sigma_{1}\otimes s_{1}$, $\Gamma_{15}=\sigma_{2}\otimes s_{1}$, $\Gamma_{23}=\sigma_{0}\otimes s_{1}$, $\Gamma_{24}=\sigma_{1}\otimes s_{2}$, $\Gamma_{25}=\sigma_{2}\otimes s_{2}$, $\Gamma_{34}=\sigma_{1}\otimes s_{3}$, $\Gamma_{35}=\sigma_{2}\otimes s_{3}$, and $\Gamma_{45}=\sigma_{3}\otimes s_{0}$. Note that, we have multiplied a factor of $\sigma_{0}\otimes is_{2}$ to each component of the basis sets. To get the final expressions for the pairing components, we have to multiply back a factor of $\sigma_{0}\otimes (-is_{2})$ to each component listed in the tables \cite{blount85,yip13,yip14,note1}. Also notice that, each basis set can be multiplied by a factor of an arbitrary linear combination of $\varphi_{0}(\mathbf{k})$ and a constant.

\begin{table}[ht] 
\caption{Basis functions for the even-parity two-dimensional representation $E_{g}$, expressed as linear combinations of products between the $\Gamma$ matrices and the symmetrized Fourier functions. The first column is the numbering of the various pairing channels. The second and third columns are separately the two components of the corresponding basis sets.} \centering
\begin{tabular}{c c c}
\hline\hline \\ [-1.5ex]
$E_{g}^{(n)}$ \hspace{1cm} & $\psi_{1}^{(n)}(\mathbf{k})(\sigma_{0}\otimes is_{2})$ \hspace{1cm} & $\psi_{2}^{(n)}(\mathbf{k})(\sigma_{0}\otimes is_{2})$ \\ [0.2ex]
\hline  \\ [-2ex]
$n=1$  \hspace{1cm} & -$I_{4}\varphi_{2}(\mathbf{k})$  \hspace{1cm} & $I_{4}\varphi_{1}(\mathbf{k})$  \\
\hline  \\ [-2ex]
$n=2$  \hspace{1cm} & -$\Gamma_{5}\varphi_{2}(\mathbf{k})$  \hspace{1cm} & $\Gamma_{5}\varphi_{1}(\mathbf{k})$  \\
\hline  \\ [-2ex]
$n=3$  \hspace{1cm} & -$\Gamma_{4}\varphi_{6}(\mathbf{k})$  \hspace{1cm} & $\Gamma_{4}\varphi_{5}(\mathbf{k})$  \\
\hline  \\ [-2ex]
$n=4$  \hspace{1cm} & $\Gamma_{2}\varphi_{3}(\mathbf{k})$  \hspace{1cm} & -$\Gamma_{1}\varphi_{3}(\mathbf{k})$  \\
\hline  \\ [-2ex]
$n=5$  \hspace{1cm} & $\Gamma_{1}\varphi_{4}(\mathbf{k})$  \hspace{1cm} & $\Gamma_{2}\varphi_{4}(\mathbf{k})$  \\
\hline  \\ [-2ex]
$n=6$  \hspace{1cm} & $\Gamma_{3}\varphi_{5}(\mathbf{k})$  \hspace{1cm} & $\Gamma_{3}\varphi_{6}(\mathbf{k})$  \\
\hline  \\ [-2ex]
$n=7$  \hspace{1cm} & $\Gamma_{2}\varphi_{5}(\mathbf{k})$+$\Gamma_{1}\varphi_{6}(\mathbf{k})$  \hspace{1cm} & -$\Gamma_{2}\varphi_{6}(\mathbf{k})$+$\Gamma_{1}\varphi_{5}(\mathbf{k})$ \\
\hline
\hline
\end{tabular}
\end{table}

\begin{table}[ht] 
\caption{Basis functions for the odd-parity two-dimensional representation $E_{u}$, expressed as linear combinations of products between the $\Gamma$ matrices and the symmetrized Fourier functions. The first column is the numbering of the various pairing channels. The second and third columns are separately the two components of the corresponding basis sets.} \centering
\begin{tabular}{c c c}
\hline\hline \\ [-1.5ex]
$E_{u}^{(n)}$ \hspace{0.5cm} & $\tilde{\psi}_{1}^{(n)}(\mathbf{k})(\sigma_{0}\otimes is_{2})$ \hspace{0.5cm} & $\tilde{\psi}_{2}^{(n)}(\mathbf{k})(\sigma_{0}\otimes is_{2})$ \\ [0.2ex]
\hline  \\ [-2ex]
$n=1$  \hspace{0.5cm} & $\Gamma_{15}$  \hspace{0.5cm} & $\Gamma_{25}$  \\
\hline  \\ [-2ex]
$n=2$  \hspace{0.5cm} & $\Gamma_{35}\varphi_{1}(\mathbf{k})$  \hspace{0.5cm} & $\Gamma_{35}\varphi_{2}(\mathbf{k})$  \\
\hline  \\ [-2ex]
$n=3$  \hspace{0.5cm} & $\Gamma_{45}\varphi_{2}(\mathbf{k})$  \hspace{0.5cm} & -$\Gamma_{45}\varphi_{1}(\mathbf{k})$  \\
\hline  \\ [-2ex]
$n=4$  \hspace{0.5cm} & $\Gamma_{15}\varphi_{2}(\mathbf{k})$+$\Gamma_{25}\varphi_{1}(\mathbf{k})$  \hspace{0.5cm} & $\Gamma_{15}\varphi_{1}(\mathbf{k})$-$\Gamma_{25}\varphi_{2}(\mathbf{k})$ \\
\hline  \\ [-2ex]
$n=5$  \hspace{0.5cm} & $\Gamma_{12}\varphi_{5}(\mathbf{k})$  \hspace{0.5cm} & $\Gamma_{12}\varphi_{6}(\mathbf{k})$  \\
\hline  \\ [-2ex]
$n=6$  \hspace{0.5cm} & $\Gamma_{34}\varphi_{5}(\mathbf{k})$  \hspace{0.5cm} & $\Gamma_{34}\varphi_{6}(\mathbf{k})$  \\
\hline  \\ [-2ex]
$n=7$  \hspace{0.5cm} & $\Gamma_{13}\varphi_{3}(\mathbf{k})$  \hspace{0.5cm} & $\Gamma_{23}\varphi_{3}(\mathbf{k})$  \\
\hline  \\ [-2ex]
$n=8$  \hspace{0.5cm} & -$\Gamma_{24}\varphi_{3}(\mathbf{k})$  \hspace{0.5cm} & $\Gamma_{14}\varphi_{3}(\mathbf{k})$  \\
\hline  \\ [-2ex]
$n=9$  \hspace{0.5cm} & $\Gamma_{23}\varphi_{4}(\mathbf{k})$  \hspace{0.5cm} & -$\Gamma_{13}\varphi_{4}(\mathbf{k})$  \\
\hline  \\ [-2ex]
$n=10$  \hspace{0.5cm} & $\Gamma_{14}\varphi_{4}(\mathbf{k})$  \hspace{0.5cm} & $\Gamma_{24}\varphi_{4}(\mathbf{k})$  \\
\hline  \\ [-2ex]
$n=11$  \hspace{0.5cm} & $-\Gamma_{13}\varphi_{5}(\mathbf{k})$+$\Gamma_{23}\varphi_{6}(\mathbf{k})$  \hspace{0.5cm} & $\Gamma_{13}\varphi_{6}(\mathbf{k})$+$\Gamma_{23}\varphi_{5}(\mathbf{k})$ \\
\hline  \\ [-2ex]
$n=12$  \hspace{0.5cm} & $\Gamma_{14}\varphi_{6}(\mathbf{k})$+$\Gamma_{24}\varphi_{5}(\mathbf{k})$  \hspace{0.5cm} & $\Gamma_{14}\varphi_{5}(\mathbf{k})$-$\Gamma_{24}\varphi_{6}(\mathbf{k})$ \\
\hline
\hline
\end{tabular}
\end{table}

The symmetry channels listed in Table III and Table IV are one central result of this work. Two features of the Tables are apparent. Firstly, only the two basis functions of $E_{u}^{(1)}$ are completely $\mathbf{k}$-independent. This is the pairing channel that has attracted the most attention as a promising candidate for the nematic pairing in Cu$_{x}$Bi$_{2}$Se$_{3}$ \cite{fu14,fu16}. Secondly, among the listed basis sets in the $E_{g}$ channel, only the leading three channels are spin-singlet in the spin-orbital basis. Among the twelve $E_{u}$ channels, only $E_{u}^{(3)}$ is spin-singlet in the spin-orbital basis.

According to a theorem by Yip and Garg, the most general pairing can be written as a linear combination of all independent basis sets of the representation \cite{yip93}. We therefore write the general expression of the pairing term in the $E_{g}$ representation as
\begin{equation}
\underline{\Delta}_{g}(\mathbf{k})=\sum\limits_{\alpha=1}^{7}\Delta_{\alpha}
[\eta_{1}\psi_{1}^{(\alpha)}(\mathbf{k})+\eta_{2}\psi_{2}^{(\alpha)}(\mathbf{k})].
\end{equation}
$(\eta_{1},\eta_{2})$ is the same vector for all seven $E_{g}$ pairing channels, so that the order parameter transforms as a well-defined vector in the subspace of the $E_{g}$ representation. Similarly, the general pairing in the $E_{u}$ representation is written as
\begin{equation}
\underline{\Delta}_{u}(\mathbf{k})=\sum\limits_{\alpha=1}^{12}\tilde{\Delta}_{\alpha} [\tilde{\eta}_{1}\tilde{\psi}_{1}^{(\alpha)}(\mathbf{k})+\tilde{\eta}_{2}\tilde{\psi}_{2}^{(\alpha)}(\mathbf{k})].
\end{equation}
Again, $(\tilde{\eta}_{1},\tilde{\eta}_{2})$ are the same vector for all twelve $E_{u}$ pairing channels. Notice that, we have restricted the pairing to the same (i.e., $E_{u}$ or $E_{g}$) irreducible representation of the symmetry group. The superconducting state with mixed even-parity and odd-parity components, which was found to stabilize under suitable circumstances \cite{wu17,wang17}, will not be considered in the present work.

Among the pairing combinations contained in Eqs.(21) and (22), we are particularly interested in those pairings that have been stabilized as the ground state in previous studies based on a microscopic or phenomenological pairing mechanism, and those that are possibly consistent with more than one key experimental consensuses on the M$_{x}$Bi$_{2}$Se$_{3}$ superconductors. The theoretical studies motivated by the recent experiments focus on pairings in the odd-parity $E_{u}$ channel. Besides $E_{u}^{(1)}$ \cite{fu14,hao17}, other $E_{u}$ pairings have been studied in previous works. Direct comparison shows that $E_{u}^{(6)}$ and $E_{u}^{(11)}$ were studied by Yuan \emph{et al} \cite{yuan17}, $E_{u}^{(6)}$ and $E_{u}^{(9)}$ (if we replace $\varphi_{4}(\mathbf{k})$ with $\varphi_{4}'(\mathbf{k})=\sin\mathbf{k}\cdot\boldsymbol{\delta}_{4}$) were studied by Chirolli \emph{et al} \cite{chirolli17}. These pairing channels exhaust the three kinds of pairings in Table I and Table II, as regards the difference between the two pairing components: For $E_{u}^{(1)}$ and $E_{u}^{(9)}$ the difference comes from the two different $\Gamma$-matrices, $\Gamma_{15}$ versus $\Gamma_{25}$ for $E_{u}^{(1)}$ and $\Gamma_{13}$ versus $\Gamma_{23}$ for $E_{u}^{(9)}$; for $E_{u}^{(6)}$ the difference between the two components comes from the $\varphi_{5}(\mathbf{k})$ and $\varphi_{6}(\mathbf{k})$ symmetry factors; for $E_{u}^{(11)}$ the distinction comes from a combination of the difference between $\Gamma_{13}$ and $\Gamma_{23}$ and the difference between $\varphi_{5}(\mathbf{k})$ and $\varphi_{6}(\mathbf{k})$.

The even-parity $E_{g}$ channels have attracted much less attention. The only theoretical paper focusing on the even-parity pairings studied the leading pairing instabilities resulting from the purely repulsive short-range electron-electron interactions \cite{hao13,note1}. The six pairings found in that paper could be identified as the six pairing components of $E_{g}^{(1)}$, $E_{g}^{(2)}$, and $E_{g}^{(3)}$. In all these three $E_{g}$ channels, the difference between the two basis components comes completely from the $\mathbf{k}$-dependent symmetry factors.


The nature of the pairing defined by Eq.(21) [Eq.(22)] depends both on the pairing strengths $\Delta_{\alpha}$ ($\tilde{\Delta}_{\alpha}$) and on the two-component vector ($\eta_{1}$, $\eta_{2}$) [($\tilde{\eta}_{1}$, $\tilde{\eta}_{2}$)]. For simplicity, we will assume in the following analysis that $\Delta_{\alpha}$ ($\alpha=1,...,6$) and $\tilde{\Delta}_{\alpha}$ ($\alpha=1,...,12$) are all real numbers. By this convention, we neglect pairings analogous to the single-component chiral pairings, like the $s+is'$, $d+id'$, and $p+ip'$ pairings \cite{uchoa07,honerkamp08,lu16}. The nonzero $\Delta_{\alpha}$ or $\tilde{\Delta}_{\alpha}$ indicate the pairing channels that contribute to the superconducting order parameter. The relative magnitudes of the $\Delta_{\alpha}$ or $\tilde{\Delta}_{\alpha}$ parameters characterize the contributions of different pairing channels. Then, depending on the ($\eta_{1}$, $\eta_{2}$) or the ($\tilde{\eta}_{1}$, $\tilde{\eta}_{2}$) vector, we may define the chirality and nematicity of the two-component superconducting order parameters \cite{fu14,venderbos16a,venderbos16b}: The pairing is nematic if at least one of $|\eta_{1}|^{2}-|\eta_{2}|^{2}$ ($|\tilde{\eta}_{1}|^{2}-|\tilde{\eta}_{2}|^{2}$) and $\eta_{1}\eta_{2}^{\ast}+\eta_{1}^{\ast}\eta_{2}$ ($\tilde{\eta}_{1}\tilde{\eta}_{2}^{\ast}+\tilde{\eta}_{1}^{\ast}\tilde{\eta}_{2}$) is nonzero. The pairing is chiral if $\eta_{1}\eta_{2}^{\ast}-\eta_{1}^{\ast}\eta_{2}\ne0$ ($\tilde{\eta}_{1}\tilde{\eta}_{2}^{\ast}-\tilde{\eta}_{1}^{\ast}\tilde{\eta}_{2}\ne0$), or equivalently $\eta_{1}/\eta_{2}$ ($\tilde{\eta}_{1}/\tilde{\eta}_{2}$) is a nonzero and finite complex number.

To understand the properties of various pairings, we turn from the spin-orbital basis to the band (pseudospin) basis \cite{yip13,yip16,zocher13,hashimoto13,nagai14,takami14,hao14}. Defining $U_{\mathbf{k}}=[|\mathbf{k},\alpha\rangle,|\mathbf{k},\beta\rangle]$ and $U_{-\mathbf{k}}=[|-\mathbf{k},\alpha\rangle,|-\mathbf{k},\beta\rangle]$, for $k_{z}>0$. A pairing wave function (i.e., superconducting order parameter) expressed as $\underline{\Delta}(\mathbf{k})$ in the original spin-orbital basis transforms to
\begin{equation}
\tilde{\underline{\Delta}}(\mathbf{k})=U^{\dagger}_{\mathbf{k}}\underline{\Delta}(\mathbf{k})U^{\ast}_{-\mathbf{k}}
\end{equation}
in the pseudospin basis \cite{hao17}. As the properties of the six symmetry factors $\varphi_{i}(\mathbf{k})$ ($i=1,...,6$) are known from Eqs.(11)-(16), the remaining task is to calculate the basis transformations for the sixteen $4\times4$ matrices in Table III and Table IV.

\subsection{The $E_{g}$ pairings}

First consider the $E_{g}$ representation. Define $I'_{4}=I_{4}(-\sigma_{0}\otimes is_{2})$ and $\Gamma'_{i}=\Gamma_{i}(-\sigma_{0}\otimes is_{2})$ ($i=1,...,5$), and define the Pauli matrices in the pseudospin basis as $\varrho_{i}$ ($i=0,1,2,3$). In terms of Eq.(23), the relevant transformations are found to be
\begin{equation}
\tilde{I}'_{4}(\mathbf{k})=-i\varrho_{2},
\end{equation}
\begin{equation}
\tilde{\Gamma}'_{1}(\mathbf{k})=-\frac{A_{0}c_{y}(\mathbf{k})}{E_{\mathbf{k}}}i\varrho_{2},
\end{equation}
\begin{equation}
\tilde{\Gamma}'_{2}(\mathbf{k})=\frac{A_{0}c_{x}(\mathbf{k})}{E_{\mathbf{k}}}i\varrho_{2},
\end{equation}
\begin{equation}
\tilde{\Gamma}'_{3}(\mathbf{k})=-\frac{R_{1}d_{1}(\mathbf{k})}{E_{\mathbf{k}}}i\varrho_{2},
\end{equation}
\begin{equation}
\tilde{\Gamma}'_{4}(\mathbf{k})=-\frac{B_{0}c_{z}(\mathbf{k})+R_{2}d_{2}(\mathbf{k})}{E_{\mathbf{k}}}i\varrho_{2},
\end{equation}
\begin{equation}
\tilde{\Gamma}'_{5}(\mathbf{k})=-\frac{M(\mathbf{k})}{E_{\mathbf{k}}}i\varrho_{2}.
\end{equation}

While only three (i.e., $E_{g}^{(1)}$, $E_{g}^{(2)}$, and $E_{g}^{(3)}$) out of the seven channels listed in Table III are spin-singlet in the original spin-orbital basis, all seven $E_{g}$ channels are pseudospin singlets in the pseudospin basis. In particular, although $E_{g}^{(7)}$ is very different from $E_{g}^{(1)}$ and $E_{g}^{(2)}$ in the spin-orbital basis, the symmetry factors of the two basis components of $E_{g}^{(7)}$ behave like $k_{x}^{2}-k_{y}^{2}$ and $-2k_{x}k_{y}$ in the band (pseudospin) basis, qualitatively the same as the corresponding basis components of $E_{g}^{(1)}$ and $E_{g}^{(2)}$, if the slight anisotropy introduced by $E_{\mathbf{k}}$ and $M(\mathbf{k})$ are neglected. In addition, $E_{g}^{(4)}$ is identical to $E_{g}^{(6)}$ in the pseudospin basis, up to a $\mathbf{k}$-independent constant factor. Finally, we point out that the seemingly different $E_{g}^{(3)}$ and $E_{g}^{(5)}$ are closely related. In fact, if we replace $\varphi_{4}(\mathbf{k})=d_{2}(\mathbf{k})$ in $E_{g}^{(5)}$ by $\varphi_{4}(\mathbf{k})+\frac{B_{0}}{R_{2}}\varphi_{4}'(\mathbf{k}) =d_{2}(\mathbf{k})+\frac{B_{0}}{R_{2}}c_{z}(\mathbf{k})$, which have the same symmetry as that of $\varphi_{4}(\mathbf{k})$ under $D_{3d}$, then $E_{g}^{(3)}$ and $E_{g}^{(5)}$ are identical in the pseudospin basis up to a constant factor.

Notice that, the six spin-singlet pairings identified in a previous study combine to $E_{g}^{(1)}$, $E_{g}^{(2)}$, and $E_{g}^{(3)}$ \cite{hao13,note1}. The two components of $E_{g}^{(3)}$ were incorrectly identified as belonging to the $E_{u}$ representation \cite{hao13,note1}, because of neglecting the different transformation properties of the pairing term compared to the model for the normal state electronic structures, as we explained in Eqs.(17)-(20).

To understand the qualitative properties of the various pairing channels, we consider each pairing channel separately. That is, we assume that only one among the seven $\Delta_{\alpha}$ parameters in Eq.(21) is nonzero. From the transformations in Eqs.(24)-(29) and the expansions in Eqs.(11)-(16), it is easy to see that all the fourteen pairing components, $\psi_{m}^{(n)}$ ($n=1,...,7$ and $m=1,2$), included in Table III have line nodes, for both spheroidal and corrugated cylindrical Fermi surfaces. If the Fermi surface is corrugated cylindrical, a chiral combination of $\psi_{1}^{(1)}$ and $\psi_{2}^{(1)}$ can give a fully-gapped bulk spectrum. The same is true for the two components of $E_{g}^{(2)}$ and the two components of $E_{g}^{(7)}$. If the Fermi surface is spheroidal, the chiral combination of the two components of $E_{g}^{(1,2,7)}$ give a bulk spectrum with two point nodes at $k_{x}=k_{y}=0$ of the Fermi surface. The remaining four pairing channels, $E_{g}^{(3)}$ to $E_{g}^{(6)}$, have line nodes for arbitrary $(\eta_{1},\eta_{2})$, for both spheroidal and corrugated cylindrical Fermi surfaces. One common set of line nodes for $E_{g}^{(3)}$ comes by setting the $B_{0}c_{z}(\mathbf{k})+R_{2}d_{2}(\mathbf{k})$ factor of Eq.(28) to zero. Six line nodes persist for $E_{g}^{(4)}$ and $E_{g}^{(6)}$, which both come from $\varphi_{3}(\mathbf{k})=d_{1}(\mathbf{k})=0$. $E_{g}^{(5)}$ also has six prevalent line nodes, which come from $\varphi_{4}(\mathbf{k})=d_{2}(\mathbf{k})=0$. For a spheroidal Fermi surface, the six line nodes from $d_{1}(\mathbf{k})=d_{2}(\mathbf{k})=0$ connect at the two points of the Fermi surface with $k_{x}=k_{y}=0$.

We can also estimate the magnitude of the superconducting gaps. For M$_{x}$Bi$_{2}$Se$_{3}$ (M is Cu, Sr, or Nd), the chemical potential $\mu>0$ lies in the conduction band. According to experiments \cite{wray10,lahoud13} and first-principles calculations \cite{zhang09,liu10}, $k_{x}a$ and $k_{y}a$ are all very small for wave vectors on the Fermi surface. The superconducting gap of a certain pairing channel can therefore be characterized in terms of its power in $ka=\sqrt{k_{x}^{2}+k_{y}^{2}}a$. For states lying on the Fermi surface, we have $E_{\mathbf{k}}+\epsilon(\mathbf{k})=\mu$. $\epsilon(\mathbf{k})$ and $M(\mathbf{k})$ vary only slightly over states on the Fermi surface, and so does $E_{\mathbf{k}}$ \cite{hao17}. As an approximation, we treat $E_{\mathbf{k}}=\mu-\epsilon(\mathbf{k})$ and $M(\mathbf{k})$ as constants. Under these conditions, we see that the gap of $E_{g}^{(1)}$ is of the order $(ka)^{2}$. $E_{g}^{(2)}$ and $E_{g}^{(7)}$ also open superconducting gaps in the order of $(ka)^{2}$, but reduced by a factor of $M(\mathbf{k})/E_{\mathbf{k}}$ and $A_{0}/E_{\mathbf{k}}$ compared to $E_{g}^{(1)}$. The superconducting gaps of $E_{g}^{(4,5,6)}$ are all in the order of $(ka)^{4}$ and are two powers smaller than $(ka)^{2}$. For $c_{z}(\mathbf{k})=0$, the superconducting gap for $E_{g}^{(3)}$ is also in the order of $(ka)^{4}$. However, the two components of $E_{g}^{(3)}$ behave more like $(k_{x}a)(k_{z}c)$ and $(k_{y}a)(k_{z}c)$, and are more efficient than $E_{g}^{(4,5,6)}$ in opening the superconducting gap. The angular dependence of the pairing amplitudes on the $k_{x}k_{y}$ plane, which was neglected in the above analysis in terms of the $ka$ factor, can be obtained from Eqs.(11)-(16).

Among the seven pairing channels in Table III, only the two components of $E_{g}^{(3)}$ can have a sign change in the pseudospin basis, when we substitute $-k_{z}$ for $k_{z}$. This implies that only $E_{g}^{(3)}$ can give SABSs on the natural $xy$ surface of the M$_{x}$Bi$_{2}$Se$_{3}$ superconductors. As regards the topological surface states of the normal phase, according to a previous theoretical study \cite{hao15}, all the pairing channels in Table III except for $E_{g}^{(3)}$ can open a gap in the topological surface states. Therefore, among all seven $E_{g}$ pairings, $E_{g}^{(3)}$ is special as regards the surface properties. The surface states for the $E_{g}^{(3)}$ pairings were studied in a previous work \cite{hao13}. On the other hand, since all seven $E_{g}$ pairings are pseudospin singlets, they are expected to have trivial isotropic electronic spin susceptibility in the $xy$ plane.

\subsection{The $E_{u}$ pairings}

We next study the pairings belonging to the $E_{u}$ representation. We define $\Gamma_{\mu\nu}'=\Gamma_{\mu\nu}(-i\sigma_{0}\otimes s_{2})$ for $\mu,\nu=1,...,5$ and $\mu<\nu$. As have been noticed in previous studies, the $E_{u}^{(1)}$ channel, which is $\mathbf{k}$-independent in the spin-orbital basis, has a complicated $\mathbf{k}$-dependence in the band (pseudospin) basis \cite{yip13,fu14,hao17}. This is generally true for all the $12$ $E_{u}$ channels listed in Table II. Besides the $E_{u}^{(1)}$ channel, only the symmetry factors of $E_{u}^{(2)}$ and $E_{u}^{(3)}$ are even functions of $\mathbf{k}$. In addition, $E_{u}^{(2)}$ is a direct product of $\Gamma_{35}$ which belongs to the $A_{1u}$ representation (Table I) and $\{\varphi_{1}(\mathbf{k}),\varphi_{2}(\mathbf{k})\}$ which belongs to the $E_{g}$ representation (Table II). $\Gamma_{35}'$ was a chief candidate of the pairing for Cu$_{x}$Bi$_{2}$Se$_{3}$ in early theoretical discussions \cite{fu10}. The relevant basis transformations for these three channels are
\begin{equation}
\tilde{\Gamma}'_{15}(\mathbf{k})=[(B_{0}c_{z}+R_{2}d_{2})\varrho_{1}
-R_{1}d_{1}\varrho_{2}-A_{0}c_{x}\varrho_{3}]\frac{i\varrho_{2}}{E_{\mathbf{k}}},
\end{equation}
\begin{equation}
\tilde{\Gamma}'_{25}(\mathbf{k})=[R_{1}d_{1}\varrho_{1}+(B_{0}c_{z}+R_{2}d_{2})\varrho_{2}
-A_{0}c_{y}\varrho_{3}]\frac{i\varrho_{2}}{E_{\mathbf{k}}},
\end{equation}
\begin{equation}
\tilde{\Gamma}'_{35}(\mathbf{k})=[A_{0}c_{x}\varrho_{1}
+A_{0}c_{y}\varrho_{2}+(B_{0}c_{z}+R_{2}d_{2})\varrho_{3}]\frac{i\varrho_{2}}{E_{\mathbf{k}}},
\end{equation}
\begin{equation}
\tilde{\Gamma}'_{45}(\mathbf{k})=[-A_{0}c_{y}\varrho_{1}+A_{0}c_{x}\varrho_{2}
-R_{1}d_{1}\varrho_{3}]\frac{i\varrho_{2}}{E_{\mathbf{k}}}.
\end{equation}
The $\mathbf{k}$-dependence of the terms in the results are suppressed to simplify the notations.
In contrast to the salient two-fold anisotropy in the spin structure factors in the two bases for $E_{u}^{(1)}$, the spin structure factors for $E_{u}^{(2)}$ and $E_{u}^{(3)}$ are fairly symmetric in the $xy$-plane.

The symmetrized Fourier functions for the remaining $E_{u}$ channels are all odd functions of $\mathbf{k}$. The properties of these pairing channels can also be understood by combining the symmetry factors and the expressions of the remaining six $\Gamma$-matrices in the pseudospin basis. By straightforward applications of Eq.(23), we get the following results:
\begin{eqnarray}
&&\tilde{\Gamma}'_{12}(\mathbf{k})=\{-[A_{0}c_{x}(B_{0}c_{z}+R_{2}d_{2})+A_{0}c_{y}R_{1}d_{1}]\varrho_{1} \notag \\
&&-[A_{0}c_{y}(B_{0}c_{z}+R_{2}d_{2})-A_{0}c_{x}R_{1}d_{1}]\varrho_{2}  \notag \\
&&-[E_{\mathbf{k}}(E_{\mathbf{k}}+M)-A_{0}^{2}c^{2}]\varrho_{3}
\}\frac{i\varrho_{2}}{E_{\mathbf{k}}(E_{\mathbf{k}}+M)},
\end{eqnarray}
\begin{eqnarray}
&&\tilde{\Gamma}'_{34}(\mathbf{k})
=\{[A_{0}c_{x}(B_{0}c_{z}+R_{2}d_{2})+A_{0}c_{y}R_{1}d_{1}]\varrho_{1} \notag \\
&&+[A_{0}c_{y}(B_{0}c_{z}+R_{2}d_{2})-A_{0}c_{x}R_{1}d_{1}]\varrho_{2}  \\
&&-[M(E_{\mathbf{k}}+M)+A_{0}^{2}c^{2}]\varrho_{3}
\}\frac{i\varrho_{2}}{E_{\mathbf{k}}(E_{\mathbf{k}}+M)}.   \notag
\end{eqnarray}
\begin{eqnarray}
&&\tilde{\Gamma}'_{13}(\mathbf{k})=\{[R_{1}d_{1}(B_{0}c_{z}+R_{2}d_{2})-A_{0}^{2}c_{x}c_{y}]\varrho_{1} \notag \\
&&+[E_{\mathbf{k}}(E_{\mathbf{k}}+M)-A_{0}^{2}c_{y}^{2}-R_{1}^{2}d_{1}^{2}]\varrho_{2}  \\
&&-[A_{0}c_{x}R_{1}d_{1}+A_{0}c_{y}(B_{0}c_{z}+R_{2}d_{2})]\varrho_{3}
\}\frac{i\varrho_{2}}{E_{\mathbf{k}}(E_{\mathbf{k}}+M)},   \notag
\end{eqnarray}
\begin{eqnarray}
&&\tilde{\Gamma}'_{23}(\mathbf{k})
=\{-[E_{\mathbf{k}}(E_{\mathbf{k}}+M)-A_{0}^{2}c_{x}^{2}-R_{1}^{2}d_{1}^{2}]\varrho_{1} \notag \\
&&+[A_{0}^{2}c_{x}c_{y}+R_{1}d_{1}(B_{0}c_{z}+R_{2}d_{2})]\varrho_{2}  \\
&&+[A_{0}c_{x}(B_{0}c_{z}+R_{2}d_{2})-A_{0}c_{y}R_{1}d_{1}]\varrho_{3}
\}\frac{i\varrho_{2}}{E_{\mathbf{k}}(E_{\mathbf{k}}+M)},   \notag
\end{eqnarray}
\begin{eqnarray}
&&\tilde{\Gamma}'_{14}(\mathbf{k})
=\{-[E_{\mathbf{k}}(E_{\mathbf{k}}+M)-A_{0}^{2}c_{y}^{2}-(B_{0}c_{z}+R_{2}d_{2})^{2}]\varrho_{1} \notag \\
&&-[A_{0}^{2}c_{x}c_{y}+R_{1}d_{1}(B_{0}c_{z}+R_{2}d_{2})]\varrho_{2}  \\
&&-[A_{0}c_{x}(B_{0}c_{z}+R_{2}d_{2})-A_{0}c_{y}R_{1}d_{1}]\varrho_{3}
\}\frac{i\varrho_{2}}{E_{\mathbf{k}}(E_{\mathbf{k}}+M)},   \notag
\end{eqnarray}
\begin{eqnarray}
&&\tilde{\Gamma}'_{24}(\mathbf{k})
=\{-[A_{0}^{2}c_{x}c_{y}-R_{1}d_{1}(B_{0}c_{z}+R_{2}d_{2})]\varrho_{1} \notag \\
&&+[A_{0}^{2}c_{x}^{2}+(B_{0}c_{z}+R_{2}d_{2})^{2}-E_{\mathbf{k}}(E_{\mathbf{k}}+M)]\varrho_{2}  \\
&&-[A_{0}c_{x}R_{1}d_{1}+A_{0}c_{y}(B_{0}c_{z}+R_{2}d_{2})]\varrho_{3}
\}\frac{i\varrho_{2}}{E_{\mathbf{k}}(E_{\mathbf{k}}+M)}.   \notag
\end{eqnarray}
Again, the $\mathbf{k}$-dependence of the terms in the results are suppressed. The expression multiplying $\varrho_{3}$ in the third line of Eq.(35) can be rewritten as
\begin{equation}
M(E_{\mathbf{k}}+M)+A_{0}^{2}c^{2} =E_{\mathbf{k}}(E_{\mathbf{k}}+M)-R_{1}^{2}d_{1}^{2}-(B_{0}c_{z}+R_{2}d_{2})^{2}. \notag
\end{equation}
The transformations in Eqs.(30)-(39) and those in Eqs.(24)-(29) are another central result of the present work.

There is an interesting relation among the six basis transformations in Eqs. (34)-(39). According to Table I, the six primed $\Gamma$-matrices separate into three pairs: $\Gamma_{12}'$ and $\Gamma_{34}'=\Gamma_{12}'\Gamma_{5}$, $\Gamma_{13}'$ and $\Gamma_{42}'=\Gamma_{13}'\Gamma_{5}$, $\Gamma_{23}'$ and $\Gamma_{14}'=\Gamma_{23}'\Gamma_{5}$. The connection between the two components of each pair is revealed by summing over the corresponding expressions in the pseudospin basis, which gives
\begin{equation}
\tilde{\Gamma}_{12}'(\mathbf{k})+\tilde{\Gamma}_{34}'(\mathbf{k})=-\frac{E_{\mathbf{k}}+M}{E_{\mathbf{k}}}\varrho_{3}(i\varrho_{2}),
\end{equation}
\begin{equation}
\tilde{\Gamma}_{23}'(\mathbf{k})+\tilde{\Gamma}_{14}'(\mathbf{k})=-\frac{E_{\mathbf{k}}+M}{E_{\mathbf{k}}}\varrho_{1}(i\varrho_{2}),
\end{equation}
\begin{equation}
\tilde{\Gamma}_{31}'(\mathbf{k})+\tilde{\Gamma}_{24}'(\mathbf{k})=-\frac{E_{\mathbf{k}}+M}{E_{\mathbf{k}}}\varrho_{2}(i\varrho_{2}).
\end{equation}
Note that $I_{4}+\Gamma_{5}=I_{4}+P$, which underlies the above combinations, is the projection operator  to the even-parity subspace in the spin-orbital basis. Eqs. (40)-(42) have a clear cyclical structure. This should be related to the definition of the pseudospin basis, which was chosen to make the even-parity component of the magnetic moment operator to transform like a proper axial vector \cite{hao17}. The combinations corresponding to the projection operator $I_{4}-\Gamma_{5}=I_{4}-P$, which projects to the odd-parity subspace, are
\begin{eqnarray}
&&\tilde{\Gamma}_{12}'(\mathbf{k})-\tilde{\Gamma}_{34}'(\mathbf{k}) \notag \\
&&=\{-2[A_{0}c_{x}(B_{0}c_{z}+R_{2}d_{2})+A_{0}c_{y}R_{1}d_{1}]\varrho_{1} \notag \\
&&-2[A_{0}c_{y}(B_{0}c_{z}+R_{2}d_{2})-A_{0}c_{x}R_{1}d_{1}]\varrho_{2}   \\
&&+[A_{0}^{2}c^{2}-R_{1}^{2}d_{1}^{2}-(B_{0}c_{z}+R_{2}d_{2})^{2}]\varrho_{3}
\}\frac{i\varrho_{2}}{E_{\mathbf{k}}(E_{\mathbf{k}}+M)}, \notag
\end{eqnarray}
\begin{eqnarray}
&&\tilde{\Gamma}_{23}'(\mathbf{k})-\tilde{\Gamma}_{14}'(\mathbf{k}) \notag \\
&&=\{[A_{0}^{2}(c_{x}^{2}-c_{y}^{2})+R_{1}^{2}d_{1}^{2}-(B_{0}c_{z}+R_{2}d_{2})^{2}]\varrho_{1} \notag \\
&&+2[A_{0}^{2}c_{x}c_{y}+R_{1}d_{1}(B_{0}c_{z}+R_{2}d_{2})]\varrho_{2}   \\
&&+2[A_{0}c_{x}(B_{0}c_{z}+R_{2}d_{2})-A_{0}c_{y}R_{1}d_{1}]\varrho_{3}
\}\frac{i\varrho_{2}}{E_{\mathbf{k}}(E_{\mathbf{k}}+M)}, \notag
\end{eqnarray}
\begin{eqnarray}
&&\tilde{\Gamma}_{31}'(\mathbf{k})-\tilde{\Gamma}_{24}'(\mathbf{k}) \notag \\
&&=\{2[A_{0}^{2}c_{x}c_{y}-R_{1}d_{1}(B_{0}c_{z}+R_{2}d_{2})]\varrho_{1} \notag \\
&&+[A_{0}^{2}(c_{y}^{2}-c_{x}^{2})+R_{1}^{2}d_{1}^{2}-(B_{0}c_{z}+R_{2}d_{2})^{2}]\varrho_{2}   \\
&&+2[A_{0}c_{x}R_{1}d_{1}+A_{0}c_{y}(B_{0}c_{z}+R_{2}d_{2})]\varrho_{3}
\}\frac{i\varrho_{2}}{E_{\mathbf{k}}(E_{\mathbf{k}}+M)}. \notag
\end{eqnarray}
Although more complex than Eqs.(40)-(42), Eqs.(43)-(45) are simpler than Eqs.(34)-(39) because the $E_{\mathbf{k}}(E_{\mathbf{k}}+M)$ factors are eliminated from the numerators.

An implication of Eqs.(40)-(45) is that, we can rehybridize four pairs of the $E_{u}$ channels in Table IV to simplify the discussions on the properties of those pairing channels. Explicitly, the four pairs include $\{E_{u}^{(5)},E_{u}^{(6)}\}$, $\{E_{u}^{(7)},E_{u}^{(8)}\}$, $\{E_{u}^{(9)},E_{u}^{(10)}\}$, and $\{E_{u}^{(11)},E_{u}^{(12)}\}$. The eight new pairing combinations are defined in Table V. The expressions in the pseudospin basis of the eight pairing channels in Table V follow directly from the definitions in Table IV and Eqs.(40)-(45), and will not be written out explicitly. In shifting from Table IV to Table V, Eq.(22) is also changed in a straightforward manner. For example, the pairing components corresponding to $E_{u}^{(5)}$ and $E_{u}^{(6)}$ are changed to $E_{u}^{(1p)}$ and $E_{u}^{(1m)}$ by
\begin{eqnarray}
&&\sum\limits_{\alpha=5}^{6}\tilde{\Delta}_{\alpha} [\tilde{\eta}_{1}\tilde{\psi}_{1}^{(\alpha)}(\mathbf{k})+\tilde{\eta}_{2}\tilde{\psi}_{2}^{(\alpha)}(\mathbf{k})] \notag \\
&&=\sum\limits_{\alpha'=1p}^{1m}\tilde{\Delta}_{\alpha'} [\tilde{\eta}_{1}\tilde{\psi}_{1}^{(\alpha')}(\mathbf{k})+\tilde{\eta}_{2}\tilde{\psi}_{2}^{(\alpha')}(\mathbf{k})],
\end{eqnarray}
where the $(\tilde{\eta}_{1},\tilde{\eta}_{2})$ vector does not change, $\tilde{\Delta}_{1p}=(\tilde{\Delta}_{5}+\tilde{\Delta}_{6})/2$ and $\tilde{\Delta}_{1m}=(\tilde{\Delta}_{5}-\tilde{\Delta}_{6})/2$. Hereafter, we consider $E_{u}^{(1)}$ to $E_{u}^{(4)}$ in Table IV and the eight pairings in Table V as the $12$ independent $E_{u}$ pairing channels.

\begin{table}[ht] 
\caption{Redefinitions of the basis functions for the eight channels of the $E_{u}$ representation in Table IV, from $E_{u}^{(5)}$ to $E_{u}^{(12)}$.} \centering
\begin{tabular}{c c c}
\hline\hline \\ [-1.5ex]
$E_{u}^{(n')}$ \hspace{0.5cm} & $\tilde{\psi}_{1}^{(n')}(\mathbf{k})$ \hspace{0.5cm} & $\tilde{\psi}_{2}^{(n')}(\mathbf{k})$ \\ [0.2ex]
\hline  \\ [-2ex]
$n'=1p$  \hspace{0.5cm} & $\tilde{\psi}_{1}^{(5)}(\mathbf{k})+\tilde{\psi}_{1}^{(6)}(\mathbf{k})$  \hspace{0.5cm} & $\tilde{\psi}_{2}^{(5)}(\mathbf{k})+\tilde{\psi}_{2}^{(6)}(\mathbf{k})$  \\
\hline  \\ [-2ex]
$n'=1m$  \hspace{0.5cm} & $\tilde{\psi}_{1}^{(5)}(\mathbf{k})-\tilde{\psi}_{1}^{(6)}(\mathbf{k})$  \hspace{0.5cm} & $\tilde{\psi}_{2}^{(5)}(\mathbf{k})-\tilde{\psi}_{2}^{(6)}(\mathbf{k})$  \\
\hline  \\ [-2ex]
$n'=2p$  \hspace{0.5cm} & $\tilde{\psi}_{1}^{(7)}(\mathbf{k})+\tilde{\psi}_{1}^{(8)}(\mathbf{k})$  \hspace{0.5cm} & $\tilde{\psi}_{2}^{(7)}(\mathbf{k})+\tilde{\psi}_{2}^{(8)}(\mathbf{k})$  \\
\hline  \\ [-2ex]
$n'=2m$  \hspace{0.5cm} & $\tilde{\psi}_{1}^{(7)}(\mathbf{k})-\tilde{\psi}_{1}^{(8)}(\mathbf{k})$  \hspace{0.5cm} & $\tilde{\psi}_{2}^{(7)}(\mathbf{k})-\tilde{\psi}_{2}^{(8)}(\mathbf{k})$  \\
\hline  \\ [-2ex]
$n'=3p$  \hspace{0.5cm} & $\tilde{\psi}_{1}^{(9)}(\mathbf{k})+\tilde{\psi}_{1}^{(10)}(\mathbf{k})$  \hspace{0.5cm} & $\tilde{\psi}_{2}^{(9)}(\mathbf{k})+\tilde{\psi}_{2}^{(10)}(\mathbf{k})$  \\
\hline  \\ [-2ex]
$n'=3m$  \hspace{0.5cm} & $\tilde{\psi}_{1}^{(9)}(\mathbf{k})-\tilde{\psi}_{1}^{(10)}(\mathbf{k})$  \hspace{0.5cm} & $\tilde{\psi}_{2}^{(9)}(\mathbf{k})-\tilde{\psi}_{2}^{(10)}(\mathbf{k})$  \\
\hline  \\ [-2ex]
$n'=4p$  \hspace{0.5cm} & $\tilde{\psi}_{1}^{(11)}(\mathbf{k})+\tilde{\psi}_{1}^{(12)}(\mathbf{k})$  \hspace{0.5cm} & $\tilde{\psi}_{2}^{(11)}(\mathbf{k})+\tilde{\psi}_{2}^{(12)}(\mathbf{k})$  \\
\hline  \\ [-2ex]
$n'=4m$  \hspace{0.5cm} & $\tilde{\psi}_{1}^{(11)}(\mathbf{k})-\tilde{\psi}_{1}^{(12)}(\mathbf{k})$  \hspace{0.5cm} & $\tilde{\psi}_{2}^{(11)}(\mathbf{k})-\tilde{\psi}_{2}^{(12)}(\mathbf{k})$  \\
\hline
\hline
\end{tabular}
\end{table}

Now we make an order of magnitude estimation over the superconducting gap amplitudes of the various pairing channels in Table IV and Table V, by taking advantage of the basis transformations of the $\Gamma$-matrices in Eqs.(30)-(45) and the expansions of the symmetry factors in Eqs.(11)-(16). As we have explained in the previous subsection, $k_{x}a$ and $k_{y}a$ are all very small for wave vectors on the Fermi surface. The pairing amplitude of a pairing channel can be characterized in terms of its power in $ka=\sqrt{k_{x}^{2}+k_{y}^{2}}a$. Also, as an approximation, we may treat $E_{\mathbf{k}}=\mu-\epsilon(\mathbf{k})$ and $E_{\mathbf{k}}+M(\mathbf{k})$ as constants. Under the above conditions, the gap of $E_{u}^{(1)}$ is in the order of $ka$, the gaps of $E_{u}^{(2,3,4)}$ are in the order of $(ka)^{3}$. While $E_{u}^{(1p)}$ and $E_{u}^{(4p)}$ open gaps in the the order of $ka$, $E_{u}^{(1m)}$ and $E_{u}^{(4m)}$ open gaps in the order of $(ka)^{3}$. $E_{u}^{(2p)}$ and $E_{u}^{(3p)}$ also open gaps in the order of $(ka)^{3}$. The gap opened by $E_{u}^{(2m)}$ and $E_{u}^{(3m)}$ are of the order $(ka)^{5}$. The dependence of the superconducting gap amplitudes on the directions of the wave vectors on the $k_{x}k_{y}$ plane can be read from the expansions in Eqs.(11)-(16). Notice that, if the Fermi surface is corrugated cylindrical, the $c_{z}(\mathbf{k})$ factor undergoes drastic variations in scanning over the Fermi surface and may dominate the variation of the pairing amplitude \cite{hao17}. The above analysis is then an estimate of the bulk pairing magnitudes on the $k_{z}=0$ and $k_{z}=\pm\pi$ planes of the BZ. From the above analysis, $E_{u}^{(1)}$, $E_{u}^{(1p)}$ and $E_{u}^{(4p)}$ are the most efficient in opening a large superconducting gap. In contrast, the $E_{u}^{(2m)}$ and $E_{u}^{(3m)}$ channels are least effective in producing a superocnducting gap, and are unlikely to be the leading pairing channels. The remaining seven $E_{u}$ pairing channels have the intermediate ability in opening a bulk superconducting gap.

The surface states on the natural $xy$ surface of superconducting M$_{x}$Bi$_{2}$Se$_{3}$ may include the SABSs and the topological surface states. The SABSs are directly related to the superconducting order parameter, and exist (do not exist) if the superconducting pairing wave function undergoes (does not undergo) a sign reversal upon the reflection changing $k_{z}$ to $-k_{z}$. By this rule, it is easy to see that only five pairing channels \emph{do not} support SABSs on the $xy$ surface, including $E_{u}^{(3)}$, $E_{u}^{(1p)}$, $E_{u}^{(2p)}$, $E_{u}^{(3p)}$, and $E_{u}^{(4p)}$. The topological surface states inherited from the normal state may or may not open a gap at the Fermi level in the superconducting phase, depending both on the nature of the bulk pairing and on the nature of the topological surface states. A full list of the pairings that open a gap in the topological surface states of Bi$_{2}$Se$_{3}$ was constructed previously \cite{hao15,hao11}. According to that list, $E_{u}^{(1)}$, $E_{u}^{(2)}$, and $E_{u}^{(4)}$ cannot gap the topological surface states. On the other hand, all the remaining pairings can at least partially gap the topological surface states. For example, $\tilde{\psi}_{1}^{(3)}$ can gap all the topological surface states at the Fermi level, except for the crossing points between the topological surface states at the Fermi level and the lines determined by $\varphi_{2}(\mathbf{k})=0$.

To infer the qualitative behaviors of the electronic spin susceptibility relevant to the Knight shift experiment, we reformulate the pairing defined by Eq.(23) in the standard expression as
\begin{equation}
\tilde{\underline{\Delta}}(\mathbf{k})=[d_{0}(\mathbf{k})\varrho_{0} +\mathbf{d}(\mathbf{k})\cdot\boldsymbol{\varrho}]i\varrho_{2},
\end{equation}
where $d_{0}(\mathbf{k})$ is the pseudospin-singlet component of the pairing, $\mathbf{d}(\mathbf{k})=(d_{1}(\mathbf{k}),d_{2}(\mathbf{k}),d_{3}(\mathbf{k}))$ is a three-component vector for the pseudospin-triplet part of the pairing, and  $\boldsymbol{\varrho}=(\varrho_{1},\varrho_{2},\varrho_{3})$. From the above results, we have $\mathbf{d}(\mathbf{k})=(0,0,0)$ for the even-parity $E_{g}$ pairings and $d_{0}(\mathbf{k})=0$ for the odd-parity $E_{u}$ pairings. The $\mathbf{d}(\mathbf{k})$ vector of an $E_{u}$ pairing is perpendicular to the spin direction of the corresponding Cooper pair \cite{leggett75,joynt02,mackenzie03,maeno12}. The spin susceptibilities show distinct behaviors depending on the relative orientation between the external magnetic field $\mathbf{H}$ and the $\mathbf{d}(\mathbf{k})$ vector \cite{hashimoto13,matano16,leggett75,joynt02,mackenzie03,maeno12}: For $\mathbf{H}\perp\mathbf{d}(\mathbf{k})$, the spin susceptibility barely changes across the superconducting transition temperature $T_{c}$; for $\mathbf{H}\parallel\mathbf{d}(\mathbf{k})$, the spin susceptibility decreases below $T_{c}$ similar to a spin-singlet superconductor. In this manner, we can understand the qualitative behaviors of the electronic spin susceptibilities for the $E_{u}$ pairing channels in Table IV and Table V, in terms of Eqs.(30)-(45). For simplicity and without losing much rigor, we can first set $R_{1}=R_{2}=0$ and work with the effective pairing in the simplified model. It is easy to see that both of the two components of the $E_{u}^{(1)}$ channel break the spin rotational symmetry in the $xy$-plane to two-fold symmetry. It is also clear that both $E_{u}^{(2)}$ and $E_{u}^{(3)}$ respect the spin rotational symmetry in the $xy$ plane. For $E_{u}^{(4)}$, while the two $\Gamma$-matrices that are contained in each channel breaks the spin rotational symmetry in the $xy$-plane, the linear combination restores this symmetry. $E_{u}^{(1p)}$ and $E_{u}^{(1m)}$ respect the spin rotational symmetry in the $xy$-plane. The two components of $E_{u}^{(2p)}$ and $E_{u}^{(3p)}$ both break the three-fold in-plane spin rotational symmetry in the $xy$-plane to two-fold symmetry. The two components of $E_{u}^{(2m)}$ and $E_{u}^{(3m)}$ only slightly break the three-fold in-plane spin rotational symmetry in the $xy$-plane. The components of $E_{u}^{(4p)}$ and $E_{u}^{(4m)}$ respect the spin rotational symmetry in the $xy$-plane.

\section{comparison to experimental results}

In this section, we first make a survey over the main experimental consensuses on the three superconductors. Then we combine the results in the previous section to infer the most probable pairings for the three superconductors. To be clear, we categorize the relevant experiments on the M$_{x}$Bi$_{2}$Se$_{3}$ (M is Cu, Sr, or Nd) superconductors into three broad classes, which separately probe the bulk spectrum, the surface spectrum (along the natural $xy$ surface parallel to the basal plane), and the magnetic properties. Since both the bulk spectrum and the surface spectrum influence the magnetic properties, this division is by no means absolute.

For the bulk spectrum, both Cu$_{x}$Bi$_{2}$Se$_{3}$ \cite{kriener11,levy13} and Sr$_{x}$Bi$_{2}$Se$_{3}$ \cite{du17nc} are reported to be fully-gapped. However, there is no solid consensus on the momentum-dependence of the band gap. For Cu$_{x}$Bi$_{2}$Se$_{3}$, while an STM experiment gives a tunneling spectrum consistent with isotropic $s$-wave pairing \cite{levy13}, a later field-angle-dependent specific heat experiment suggests an energy gap structure with salient two-fold symmetry in the $k_{x}k_{y}$ plane \cite{yonezawa17}. For Sr$_{x}$Bi$_{2}$Se$_{3}$, an STM experiment implies an anisotropic $s$-wave pairing \cite{du17nc}. Later, an experiment shows with resistivity and Laue diffraction measurements that the energy gap structure of Sr$_{x}$Bi$_{2}$Se$_{3}$ also has two-fold symmetry in the $k_{x}k_{y}$ plane \cite{du17}. For Nd$_{x}$Bi$_{2}$Se$_{3}$, an experiment finds evidence for point nodes in its bulk spectrum \cite{smylie16}.


For the surface spectrum, the situation is rather confusing, for all three superconductors. While several early experiments, in particular those based on point-contact spectroscopy, claim to have found evidence of SABSs for Cu$_{x}$Bi$_{2}$Se$_{3}$ \cite{sasaki11,kirzhner12,chen12}, later experiments denied the existence of SABSs \cite{levy13,peng13}. Similar confusion exists for Sr$_{x}$Bi$_{2}$Se$_{3}$. An experiment infers the existence of surface states in Sr$_{x}$Bi$_{2}$Se$_{3}$, through the Shubnikov-de Haas oscillations \cite{liu15}. However, an STM/STS experiment does not see any in-gap states \cite{du17nc}. For superconducting Nd$_{x}$Bi$_{2}$Se$_{3}$, the angle-resolved photoemission spectroscopy experiment by Qiu \emph{et al} shows that the topological surface states in the normal state is preserved in the superconducting state \cite{qiu15}. But it is unclear whether or not the topological surface states are gapped at the Fermi level, and whether or not there are SABSs on the $xy$ surface of Nd$_{x}$Bi$_{2}$Se$_{3}$.

For the magnetic properties, Nd$_{x}$Bi$_{2}$Se$_{3}$ was reported to have spontaneous magnetization in the superconducting phase \cite{qiu15}, which seems to be related to the magnetism of the Nd dopants. The Cu$_{x}$Bi$_{2}$Se$_{3}$ and Sr$_{x}$Bi$_{2}$Se$_{3}$ superconductors are usually considered as nonmagnetic in the absence of external magnetic fields \cite{qiu15}. On the other hand, for all three superconductors, there are magnetism-related experiments indicating that the three-fold rotational symmetry in the normal phase is broken down to two-fold rotational symmetry in the superconducting phase. The Knight shift experiment by Matano \emph{et al} shows that the electronic spin susceptibilities of Cu$_{x}$Bi$_{2}$Se$_{3}$ is two-fold symmetric in the basal plane and is invariant for an out-of-plane magnetic field \cite{matano16}. Another magnetic measurement is the upper critical field, $B_{c2}$. By varying the direction of the magnetic field in the basal plane of the superconductors, a two-fold symmetry in the upper critical field is observed for Cu$_{x}$Bi$_{2}$Se$_{3}$ \cite{yonezawa17}, for Sr$_{x}$Bi$_{2}$Se$_{3}$ \cite{pan16}, and also for Nd$_{x}$Bi$_{2}$Se$_{3}$ \cite{shen17}. For Nd$_{x}$Bi$_{2}$Se$_{3}$, the two-fold rotational symmetry in the basal plane is also confirmed in a torque magnetometry measurement \cite{asaba17}.

In the light of the above survey, the most well-established experimental feature is the two-fold rotational symmetry in the basal plane, for all three superconductors. This two-fold symmetry has two incarnations, (1) in the electronic spin susceptibility, and (2) in the momentum-dependence of the superconducting gap amplitude. As we explain below, these two aspects of the two-fold symmetry are independent of each other.

The nematic combination of two components of $E_{u}^{(1)}$ is the most well-known candidate for the two-fold symmetry in the electronic spin susceptibility \cite{fu14,fu16,matano16}. $E_{u}^{(2p)}$ and $E_{u}^{(3p)}$ can also lead to two-fold symmetry in the spin susceptibility in the $xy$ plane. Because for $E_{u}^{(2p)}$ and $E_{u}^{(3p)}$ the $\mathbf{d}(\mathbf{k})$ vectors in the pseudospin basis do not have the third component, as shown in Eqs.(41) and (42), the corresponding spin susceptibility will keep invariant for an out-of-plane magnetic field. On the other hand, the $\mathbf{d}(\mathbf{k})$ vectors for the two $E_{u}^{(1)}$ pairing components, as shown in Eqs.(30) and (31), have finite third components. The spin susceptibility for a pairing in the $E_{u}^{(1)}$ channel should therefore decrease slightly for an out-of-plane magnetic field \cite{hashimoto13}. In this respect, $E_{u}^{(2p)}$ and $E_{u}^{(3p)}$ fit the Knight shift experiment better than the $E_{u}^{(1)}$ pairing \cite{matano16}.

The two-fold symmetry in the $\mathbf{k}$-dependence of the superconducting gap amplitudes could be explained by the nematic realization of $E_{u}^{(1)}$ or $E_{g}^{(3)}$. $E_{u}^{(2m)}$ and $E_{u}^{(3m)}$ also result in two-fold symmetry in the $k_{x}k_{y}$ plane. But the $(ka)^{5}$ dependence of their pairing amplitude implies that they are unlikely the leading pairing instability. The remaining channels in Table III and Table IV do not give clear two-fold symmetry in the superconducting gap amplitude in the $k_{x}k_{y}$ plane to account for the experiments. Therefore, as regards the two-fold symmetry in the basal plane, the $E_{u}^{(1)}$ channel is the only channel that naturally account for both of the two aspects of the two-fold rotational symmetry. On the other hand, each incarnation of the two-fold symmetry has alternative realizations other than $E_{u}^{(1)}$.

Now we consider further constraints imposed by the nodal structures of the bulk spectra. For both Cu$_{x}$Bi$_{2}$Se$_{3}$ \cite{levy13,kriener11} and Sr$_{x}$Bi$_{2}$Se$_{3}$ \cite{du17nc}, the bulk spectra are fully gapped according to relevant experiments. This means the absence of true or approximate nodes in the bulk spectrum. $\tilde{\psi}^{(1)}_{2}(\mathbf{k})$ of $E_{u}^{(1)}$ is the only pairing component in Table III to Table V that has approximate (point) nodes, for which the gap minima is two to three orders of magnitude smaller than the gap maxima \cite{hao17}. The two components of $E_{u}^{(4p)}$ individually opens a full gap on the Fermi surface if we have a corrugated cylindrical Fermi surface, with the size of the gap scaling as $ka$ for wave vectors on the Fermi surface. For spheroidal Fermi surfaces, both of the two components of $E_{u}^{(4p)}$ give two point nodes at $k_{x}=k_{y}=0$. All the remaining individual pairing components in Table III to Table V (including $\tilde{\psi}^{(1)}_{1}(\mathbf{k})$ of $E_{u}^{(1)}$) give true nodes to the bulk spectrum, for both spheroidal and corrugated cylindrical Fermi surfaces.

For several of the pairing channels other than $E_{u}^{(4p)}$, we can also get a fully-gapped bulk spectrum by forming chiral combinations of the two pairing components, for suitable Fermi surface topology. These channels include $E_{g}^{(1,2,7)}$ of the $E_{g}$ representation, and $E_{u}^{(2,3,1p)}$ of the $E_{u}$ representation. When the Fermi surface is a corrugated cylinder, the chiral combinations (e.g., $\eta_{1}=1$ and $\eta_{2}=i$, or $\tilde{\eta}_{1}=1$ and $\tilde{\eta}_{2}=i$) of the two components of $E_{g}^{(1,2,7)}$ or $E_{u}^{(2,3,1p)}$ all lead to a fully-gapped bulk quasiparticle spectrum. For a chiral \emph{and} nematic combination (e.g., $\eta_{1}=0.5$ and $\eta_{2}=i$, or $\tilde{\eta}_{1}=0.5$ and $\tilde{\eta}_{2}=i$) of the two components of $E_{g}^{(1,2,7)}$ or $E_{u}^{(2,3,1p)}$, the bulk quasiparticle spectrum are also fully gapped, if we have a corrugated cylindrical Fermi surface. When the Fermi surface is a spheroid, these chiral or chiral and nematic pairings have two point nodes at the $k_{x}=k_{y}=0$ points of the Fermi surface. None of the above pairing channels, including $E_{g}^{(1,2,7)}$ and $E_{u}^{(2,3,1p,4p)}$, can account for the two-fold symmetries observed in the experiments.

The time-reversal symmetry breaking combinations of the two components of $E_{u}^{(1)}$ have more complicated behaviors. For a spheroidal Fermi surface, the purely chiral combinations of the two components of $E_{u}^{(1)}$ have two point nodes at the $k_{x}=k_{y}=0$ points of the Fermi surface \cite{venderbos16b,yuan17}. These point nodes disappear both when we consider a chiral and nematic combination of the two components, and when the Fermi surface turns to a corrugated cylinder. On the other hand, all these various time-reversal symmetry breaking pairing combinations within the $E_{u}^{(1)}$ channel support SABSs on the $xy$ surface and keep the topological surface states at the Fermi level ungapped.

Summing up the above discussions, for Cu$_{x}$Bi$_{2}$Se$_{3}$ and Sr$_{x}$Bi$_{2}$Se$_{3}$, there is not a single pairing channel that can simultaneously explain the two-fold symmetric and fully-gapped bulk spectrum, with a time-reversal symmetric pairing combination. For Nd$_{x}$Bi$_{2}$Se$_{3}$, while the $E_{u}^{(1)}$ channel can explain the two-fold symmetry and bulk spectrum with (real or approximate) point nodes, the complexity of the Fermi surface of this compound complicates the comparison \cite{lawson16}. Focusing on Cu$_{x}$Bi$_{2}$Se$_{3}$ and Sr$_{x}$Bi$_{2}$Se$_{3}$, an implication of the above comparison is that we have to consider more than one pairing channels to simultaneously account for the fully-gapped bulk spectrum and the two-fold rotational symmetry in the basal plane. In other words, the pairing has to be \emph{multichannel}, in addition to having two components. This is the major conclusion of our comparison. What follows we explore various possible multichannel pairing combinations. We relax the constraint of time-reversal symmetry and explore all the possible pairing combinations that give both the two-fold symmetry in the basal plane and the fully-gapped bulk quasiparticle spectrum.

To get a fully-gapped bulk spectrum with two-fold symmetry in the superconducting gap amplitude, in terms of a multichannel pairing in the $E_{g}$ representation, the $(\eta_{1},\eta_{2})$ vector in Eq.(21) must be simultaneously chiral and nematic. The chirality and nematicity of the pairing are separately responsible for the fully-gapped bulk spectrum and the two-fold symmetry in the $k_{x}k_{y}$ plane. The full gap may be achieved by a chiral combination of $E_{g}^{(1)}$, or $E_{g}^{(2)}$, or $E_{g}^{(7)}$, under the premise that the Fermi surface is a corrugated cylinder. Including a finite contribution of $E_{g}^{(3)}$, the nematicity of the pairing can account for the two-fold symmetry in the gap amplitude. The fully-gapped bulk spectrum requires the strength of the $E_{g}^{(1)}$ (or $E_{g}^{(2)}$, or $E_{g}^{(7)}$) channel to be larger than the strength of the $E_{g}^{(3)}$ channel, namely $\Delta_{1}$ (or $\Delta_{2}$, or $\Delta_{7}$) is sufficiently large compared to $\Delta_{3}$. In this case, the superconducting order parameter (the wave function of the Cooper pairs) does not undergo a sign reversal upon scattering off the $xy$ surface, implying the absence of SABSs. In addition, the chiral combination of the dominant $E_{g}^{(1)}$ (or $E_{g}^{(2)}$, or $E_{g}^{(7)}$) pairing component fully-gaps the topological surface states. Therefore, there are no low-energy in-gap states on the $xy$ surface for this chiral and nematic pairing in the $E_{g}$ representation.

If we impose further the constraint of two-fold symmetry in the electronic spin susceptibility \cite{matano16}, we have to consider the $E_{u}$ channels. A combination of $E_{u}^{(1)}$ (or $E_{u}^{(2p)}$, or $E_{u}^{(3p)}$) and $E_{u}^{(4p)}$ is possible to give a purely nematic pairing that has a fully-gapped bulk spectrum, if $\tilde{\Delta}_{4p}$ is large compared to $\tilde{\Delta}_{1}$ (or $\tilde{\Delta}_{2p}$, or $\tilde{\Delta}_{3p}$). Let us define (I) the combination of $E_{u}^{(1)}$ with $E_{u}^{(4p)}$, (II) the combination of $E_{u}^{(2p)}$ with $E_{u}^{(4p)}$, and (III) the combination of $E_{u}^{(3p)}$ with $E_{u}^{(4p)}$. For the Knight shift experiment, the combinations II and III explain the invariant $c$-axis spin susceptibility better than combination I. In addition, they can also explain more naturally the absence of SABSs and the gapped topological surface states on the $xy$ surfaces of the superconductor. For these experimental features, combinations II and III are better alternatives to combination I. On the other hand, if we consider the two-fold symmetry in the superconducting gap amplitude, combination I explains it naturally, whereas combinations II and III do not lead to salient two-fold symmetry in the gap amplitude. Overall, to explain at least qualitatively the key experimental features of Cu$_{x}$Bi$_{2}$Se$_{3}$ with a nematic and time-reversal symmetric pairing, we have to consider a pairing consisting mainly $E_{u}^{(1)}$ and $E_{u}^{(4p)}$, and possibly supplemented by $E_{u}^{(2p)}$ and (or) $E_{u}^{(3p)}$. Finally, following the same analysis as that for the above chiral and nematic $E_{g}$ pairing, there are no low-energy in-gap states on the $xy$ surface. Explicitly, because $\tilde{\Delta}_{4p}$ is assumed to be large compared to $\tilde{\Delta}_{1}$ (or $\tilde{\Delta}_{2p}$, or $\tilde{\Delta}_{3p}$), the superconducting order parameter does not undergo sign change upon the substitution of $-k_{z}$ for $k_{z}$, so that there are no SABSs on the $xy$ surface. In addition, since $\tilde{\Delta}_{4p}$ gaps the topological surface states at the Fermi level, there are no topological surface states that may contribute to the low-energy in-gap states.

For other $E_{u}$ combinations, we again have to consider a chiral and nematic combination to account for the fully-gapped bulk spectrum and the two-fold in-plane rotational symmetry at the same time, similar to the $E_{g}$ case. If both the superconducting gap amplitudes and the electronic spin susceptibilities are required to be two-fold symmetric in the basal plane, we can choose the chiral and nematic combinations of $E_{u}^{(1)}$ with $E_{u}^{(2,3)}$. If only the electronic spin susceptibilities are required to be two-fold symmetric, the chiral and nematic combinations of $E_{u}^{(2p,3p)}$ with $E_{u}^{(2,3)}$ are eligible. Again, the pairing amplitudes for $E_{u}^{(2,3)}$, which give the fully-gapped bulk spectrum, should be larger than the pairing amplitudes for $E_{u}^{(1)}$. Because the dominant third component of the $\mathbf{d}(\mathbf{k})$ vector for $E_{u}^{(2)}$ is proportional to $B_{0}c_{z}+R_{2}d_{2}\simeq B_{0}c_{z}$, and both $E_{u}^{(1)}$ and $E_{u}^{(2)}$ not gap the topological surface states at the Fermi level, the chiral and nematic combination of $E_{u}^{(1)}$ and $E_{u}^{(2)}$ have SABSs and ungapped topological surface states on the $xy$ surface. The chiral and nematic combination of $E_{u}^{(1)}$ and $E_{u}^{(3)}$, with the magnitude of $\tilde{\Delta}_{3}$ much larger than the magnitude of $\tilde{\Delta}_{1}$ to ensure a fully-gapped bulk spectrum, does not have SABSs and ungapped topological surface states on the $xy$ surface.

A salient feature of the above multichannel pairings is the ubiquity of the chiral and nematic pairing combinations, in both the $E_{g}$ and the $E_{u}$ representations. Besides the broken in-plane rotational symmetry, the chiral character of the pairing implies that it can show typical signatures in experiments such as muon spin relaxation and optical Kerr effect which probe the broken time-reversal symmetry \cite{luke98,spielman92,xia06}. The chiral and nematic $E_{u}$ pairings are also nonunitary, which can be verified by proving the nonvanishing of the vector product $\mathbf{d}(\mathbf{k})\times\mathbf{d}^{\ast}(\mathbf{k})\neq0$ for general $\mathbf{k}$ \cite{ohmi96,hillier12}. Nonunitary pairings, as a special case of pairings with broken time-reversal symmetry, are known to have a spontaneous moment from the Cooper pairs \cite{ohmi96,hillier12}. Also note that, the chiral or chiral and nematic pairing combinations of the two components of $E_{u}^{(1)}$ are all nonunitary.

Another common character of the above pairing combinations is the requirement of a corrugated cylindrical Fermi surface, to have a fully-gapped bulk spectrum. The undoped Bi$_{2}$Se$_{3}$ is known to have a spheroidal Fermi surface. As the concentration of the dopants increases, it is natural to expect a continuous evolution of the Fermi surface from spheroidal to corrugated cylindrical. Alongside, the nodal structure of the pairing may also change. While the experiment of Lahoud \emph{et al} shows that the superconducting Cu$_{x}$Bi$_{2}$Se$_{3}$ has a corrugated cylindrical Fermi surface in the normal phase \cite{lahoud13}, there is presently no systematic studies on the evolution of the Fermi surface for any of the three superconductors.

Finally, it is interesting to notice that several pairing channels in Table III to Table V exhibit prominent four-fold rotational symmetry, including the $E_{g}^{(1,2,7)}$ channels of the $E_{g}$ representation and the $E_{u}^{(2,3)}$ channels of the $E_{u}$ representation, through the $\mathbf{k}$-dependence of the $\varphi_{1}(\mathbf{k})$ and $\varphi_{2}(\mathbf{k})$ symmetry factors. The four-fold symmetry also breaks the three-fold rotational symmetry of the underlying crystal lattice in the normal phase. Several of the chiral and nematic pairing combinations proposed above contain these pairing channels. They are therefore expected to exhibit some characters of the four-fold symmetry. On the experimental side, a four-fold symmetric component in the superconducting gap amplitude was indeed observed by Du \emph{et al} for Sr$_{x}$Bi$_{2}$Se$_{3}$ \cite{du17nc}. These chiral and nematic pairings are however inconsistent with the general belief that the superconducting state of Sr$_{x}$Bi$_{2}$Se$_{3}$ preserves the time-reversal symmetry.

\section{implications for future experiments}

From the analysis of the previous section, the available experiments have imposed stringent constraints on the true pairing symmetries of the M$_{x}$Bi$_{2}$Se$_{3}$ (M is Cu, Sr, or Nd) superconductors. In particular, they point to the multichannel two-component pairings. These experiments, on the other hand, do not provide sufficient and consistent information to unambiguously identify the true pairing symmetry of these superconductors. In particular, for each of the three superconductors, the parameter $x$ defines a series of superconductors which might have qualitatively different Fermi surface topology in the normal state and different pairing symmetries in the superconducting state. This doping dependence has not been systematically investigated for any of three superconductors, although the phase diagram of Cu$_{x}$Bi$_{2}$Se$_{3}$ exists \cite{kriener11b}. It is therefore highly desirable to make a systematic experimental study on each member of the series with a complementary set of experimental tools.

Based on the survey over the experimental consensuses and the comparison with the possible pairing channels, the relevant properties and the experiments that may be performed to probe them include: (1) The evolution of the Fermi surface with the doping concentration $x$. The magnetic oscillation experiments, including the Shubnikov-de Haas oscillation and the de Haas-van Alphen effect, can probe the Fermi surface contour in the normal phase \cite{lahoud13,lawson16,lawson12}. From the discussions of the above section, the geometry of the Fermi surface significantly influences the bulk quasiparticle spectrum of the superconducting state. It is highly desirable to determine the systematic evolution of the Fermi surface as the doping concentration $x$ increases. In addition, the angle-resolved photoemission spectroscopy (ARPES), which may probe the continuous evolution of the topological surface states with $x$ \cite{wray10,lahoud13,tanaka12}, supplements the magnetic oscillation experiment which gives the bulk Fermi surface. (2) The bulk quasiparticle spectrum of the superconducting state, fully gapped or not. For each doping concentration that turns the material to a superconductor, we may determine whether the superconducting state is fully gapped or not by combining the scanning tunneling spectroscopy (STS) measurements \cite{levy13} and the measurements of the zero-field bulk specific heat \cite{kriener11}. The STS experiment, besides probing the bulk quasiparticle spectrum, probes also the in-gap states on the surface of the superconductors \cite{levy13,du17nc}. The spin-lattice relaxation rates \cite{nagai15} and penetration depth \cite{smylie16} can also be used to probe the nodal structures of the bulk superconducting spectrum. (3) The momentum-dependence of the superconducting gap amplitude. This property on one hand refines the understanding obtained from the above step, on the other hand shows the presence or not of the anisotropy in the superconducting gap amplitudes. The relevant experiments include the field-angle-dependent specific heat experiments \cite{yonezawa17,willa18}, the field-angle-dependent upper critical magnetic field measurements \cite{yonezawa17,pan16,shen17}, the field-angle-dependent resistivity measurements \cite{du17}, and field-angle dependent STS \cite{tao18}. Besides the above experiments for the magnitude of the superconducting gap, the phase of the superconducting gap can be probed with the orientation-dependent Josephson junctions \cite{tanaka97}. (4) The electronic spin susceptibility, which is the most direct diagnostic tool for the structure of the $\mathbf{d}$-vectors of the pseudospin-triplet pairings. The major relevant experiment is the Knight shift measurement \cite{matano16}. (5) The persistence or not of the time-reversal symmetry in the superconducting state. The time-reversal symmetry breaking of the superconductors can be probed by several techniques, such as the muon spin resonance \cite{luke98}, optical Kerr effect \cite{spielman92,xia06}, and the Josephson effect \cite{maeno12}. The zero-field Hall effect was also used by Qiu \emph{et al} to probe the broken time-reversal symmetry in Nd$_{x}$Bi$_{2}$Se$_{3}$ \cite{qiu15}. (6) The presence or not of spontaneous magnetization. The above chiral and nematic $E_{u}$ pairings are mostly nonunitary and should lead to spontaneous magnetization. The experiment of Qiu \emph{et al} reporting the spontaneous magnetization of Nd$_{x}$Bi$_{2}$Se$_{3}$ was based on the field-dependent \emph{dc} magnetization, which is a measurement of the global magnetization \cite{qiu15}. It is desirable to study the spontaneous magnetization with a local measurement, like the polarized neutron diffraction experiments \cite{maeno12}.

For each member of the three series of superconductors, a systematic study of the above experiments should be able to identify the genuine pairing symmetry. From the breadth of the experimental tools involved, extensive experimental collaborations by sharing the same high-quality samples are highly desirable.

Besides the experiments listed above, other experiments which may probe further implications of the candidate pairings would also be of great potential interest. For example, as a natural consequence of the two-component nature of the pairing, the domain structure should exist in the superconducting state. Although most of the multichannel two-component pairings inferred from the analysis of the previous section do not support SABSs on the $xy$ surface, all the $E_{u}$ pairings should give ABSs as domain wall states for domain walls parallel to the $z$ axis, because the superconducting order parameters are odd functions of $k_{x}$ and $k_{y}$ in the pseudospin basis. Evidence of domains was reported in the experiment of Yonezawa \emph{et al} \cite{yonezawa17}. However, it is unclear whether or not there are nontrivial domain wall states. Further experiments like those performed for Sr$_{2}$RuO$_{4}$ \cite{kidwingira06} and superfluid $^{3}$He-$A$ \cite{kasai18} are highly desirable. As another example, for nonunitary pairings in the $E_{u}$ representation, there should be collective modes associated with the oscillation of the magnetization of the Cooper pairs \cite{ohmi96}. It would be interesting to detect this collective mode via experiments like electron spin resonance or ultrasound attenuation \cite{ohmi96}.

\section{summary}

Starting from a tight-binding model for the normal state electronic structures, we have constructed the full lists of two-component pairings for M$_{x}$Bi$_{2}$Se$_{3}$ (M is Cu, Sr, or Nd). We then transform the pairings to the pseudospin basis, based on which we study their qualitative properties. Comparisons to the main experimental consensuses on these superconductors show that the true pairing symmetry for them has to be multichannel, in addition to having two components. Besides a time-reversal symmetric nematic pairing belonging to the $E_{u}$ representation, we identify chiral and nematic pairings in both the $E_{g}$ and the $E_{u}$ representations. However, for all three superconductors, the existing experiments are insufficient to unambiguously determine the nature of the superconducting state. In particular, the studies on the dependence of the Fermi surface and the superconducting properties on the doping concentration $x$ are inadequate for all three superconductors. A complementary set of experiments is suggested to identify unambiguously the genuine pairing symmetries of the three superconducting electron-doped Bi$_{2}$Se$_{3}$.

\begin{acknowledgments}
L.H. thanks Ting-Kuo Lee, Sungkit Yip, and Jun Wang for helpful discussions. L.H. was supported by NSFC.11204035. C.S.T. was supported by the Robert A. Welch Foundation under Grant No. E-1146 and the Texas Center for Superconductivity at the University of Houston.
\end{acknowledgments}\index{}



\begin{references}
\bibitem{note1} In a previous work \cite{hao13}, because of not incorporating the extra factor of $\sigma_{0}\otimes(-is_{2})$, $E_{g}^{(3)}$ was mistakenly categorized as belonging to the $E_{u}$ representation.
\end{references}
\end{document}